\documentclass[11pt]{article}
   
\usepackage{epsfig}
\usepackage{amsfonts,amssymb}

\renewcommand{\theequation}{\arabic{section}.\arabic{equation}}

\newcommand{\half}{\ensuremath{\frac{1}{2}}}

\def\u{\upsilon}

\newcommand{\be}{\begin{equation}}
\newcommand{\ee}{\end{equation}}

\newcommand{\ba}{\begin{eqnarray}}
\newcommand{\ea}{\end{eqnarray}}

\newcommand{\ns}{\normalsize}

\begin{document}


\begin{titlepage}

\title{
   \hfill{\ns hep-th/0601230\\}
   \vskip 2cm
   {\Large\bf The dynamics of coset dimensional reduction.}
\\[0.5cm]}
   \setcounter{footnote}{0}
\author{
{\ns\large 
  \setcounter{footnote}{3}
  Josef L. P. Karthauser$^1$\footnote{email: jlk23@sussex.ac.uk}~, 
  P. M. Saffin$^{1,2}$\footnote{email: paul.saffin@nottingham.ac.uk}}
\\[0.5cm]
   $^1${\it\ns Department of Physics and Astronomy, University of Sussex}\\
   {\ns Falmer, Brighton, BN1 9QJ, UK.} \\[0.2em] 
   $^2${\it\ns School of Physics and Astronomy, University of Nottingham}\\
   {\ns University Park, Nottingham, NG7 2RD, UK.}
}
\date{}

\maketitle

\begin{abstract}\noindent
The evolution of multiple scalar fields in cosmology has been much studied, particularly
when the potential is formed from a series of exponentials. For a certain subclass of
such systems it is possible to get ``assisted`` behaviour, where the presence of multiple
terms in the potential effectively makes it shallower than the individual terms indicate. 
It is also known that when compactifying
on coset spaces one can achieve a consistent truncation to an effective theory which contains
many exponential terms, however, if there are too many exponentials then exact scaling
solutions do not exist.
In this paper we study the potentials arising from such compactifications
of eleven dimensional supergravity
and analyse the regions of parameter space which could lead to scaling behaviour.
\end{abstract}

\end{titlepage}

\section{Introduction}
\label{sec:introduction}

Extra dimensions are ubiquitous in unified models of gravity and
particles, dating from the early work of Kaluza and of Klein
to the modern ideas of string and M-theory. This raises the question
of what to do with the extra dimensions? In a fully dynamical
model we expect that at some time these extra dimensions should
evolve, possibly having some impact on the Universe we observe today.
In the context of dimensional reduction, where the extra dimensions
take the form of a small, compact manifold, the dynamics of the internal
space typically manifests itself through the dynamics of scalar fields
in an effective theory.
Such scalar fields are often termed moduli and their values describe
the shape and size of the compact space. The evolution of scalar fields
in cosmology is a well established area of study, finding applications in
inflationary model building, quintessence and many others. Indeed, cosmologists
seem able to solve most problems with the introduction of a new scalar field.
A particularly attractive system of scalars is the one where the
potential is formed from a series of exponential terms, such a system
can be phrased in the form of an autonomous dynamical system whose critical
points allow for a simplified description of the evolution
\cite{Copeland:1997et,Collinucci:2004iw}.
Fortunately such a nice description does not go to waste as exponential potentials are
common in unified gravity models with exponentials coming from 
dimensional reduction, gaugino condensation and instanton corrections
\cite{Derendinger:1985kk,Dine:1985rz,Witten:1996bn,Moore:2000fs}.

The critical points of the autonomous system formed in cosmology using
scalars with exponential potentials reveals that {\it tracking solutions} are
possible, and indicate that {\it assisted behaviour} can occur. By tracking
solution we mean that the scalars evolve in such a way that their energy
density remains at a constant fraction of the total energy density of the Universe,
with the rest of the energy density being composed typically of a barotropic
fluid. 
One may also have {\it scaling} behaviour, where the energy density of the field
evolves in proportion to $H^2$ without reference to other matter.
Assisted behaviour refers to the behaviour of multiple scalars, where their
combined effect can be such that the Universe expands more rapidly than one
may naively expect by looking at the individual terms in the potential.

In this paper we aim to study dimensional reduction on homogeneous manifolds, namely
those formed as a coset of compact Lie groups G/H. In order to be concrete we shall take
eleven dimensional supergravity to be our starting point, whose bosonic field
content comprises a metric and a three-form potential. As we are aiming for a four
dimensional effective theory we shall reduce on the seven dimensional cosets
which have been classified in \cite{Castellani:1983yg}. Such coset reductions are
familiar in the supergravity literature, where the supersymmetric stationary points
are well known. Here however we concern ourselves with the full effective potential
and the dynamics that follow, paying particular attention to the regimes which 
lead to tracking or scaling behaviour.

We shall structure the paper by first introducing the bosonic action of eleven dimensional
supergravity, along with the ansatz for dimensional reduction. Having done that we
shall produce the effective potentials for all the 7D cosets classified in \cite{Castellani:1983yg}.
With these in place we are able, after a brief introduction to scaling solutions, to
study the scaling properties of supergravity reduced on coset manifolds. Throughout the
paper we shall be using technical results for dimensional reduction and for cosets, 
the details of which can be found in the appendices.
We end the paper with our concluding remarks, and comments for future work.

\section{The model}
\label{sec:action}
If the Universe is fundamentally described by a higher dimensional theory then there
must be some mechanism whereby four spacetime dimensions are picked out at the
expense of others. One such mechanism is the famous Kaluza-Klein procedure, utilising
a compact space whose size sets the mass scale for a tower of massive
modes\cite{Forgacs:1978xy,Forgacs:1978xz}.  These
modes should be at a high enough scale that we have not yet observed them, leaving a 4D
low energy theory effective for the moduli fields. The underlying theory allows these moduli fields to evolve
and, as scalar fields, they are a natural source for the myriad scalars which cosmologists
require to solve various problems; the difficulty lies in finding an internal space which
has the requisite properties. The simplest type of internal manifold is
a torus \cite{Lidsey:1999mc}, however one may also
consider the internal space to be a group manifold
\cite{dewit:1963,Biswas:2004pd},
an Einstein manifold, a direct product of Einstein manifolds
\cite{Chen:2003dc}, or products of twisted manifolds
\cite{Neupane:2005nb,Neupane:2005ms}.

Here however we consider another case, namely reducing on
homogeneous manifolds described as coset spaces.
These are particularly attractive because they maintain the useful properties of group
manifolds while not being as restrictive, and they do have a long history of use in Kaluza-Klein
models, with their structure providing non-Abelian gauge groups via the Killing vectors
living on the coset \cite{Castellani:1983yg,Duff:1986hr}. 
As a way of connecting
eleven dimensional supergravity to standard model physics in four dimensions
it was pointed out in \cite{Witten:1981me} that seven is the minimum number of dimensions
for a homogeneous manifold invariant under the action of SU(3)$\otimes$SU(2)$\otimes$U(1).
Although there is a large body of work regarding cosets as internal
manifolds, the existing literature
is mostly concerned with finding stationary points of the effective potential, corresponding
to Einstein metrics on the coset space, rather than studying the dynamics of evolving
moduli. In this paper we shall seek to resolve this starting with a review of
the framework of cosets and then applying it to the study of the
evolution of coset spaces in cosmology, concentrating on eleven dimensional supergravity
as the higher dimensional theory \cite{Cremmer:1978km}. Using
the conventions of \cite{Gauntlett:2002fz} the bosonic action for the
theory is
\ba
\hat S&=&\frac{1}{2\kappa_{11}^2}\int\left[\star\hat{\cal R}-\half \star F\wedge F-\frac{1}{6}C\wedge F\wedge F\right],
\ea
where $\hat{\cal R}$ is the 11D Ricci scalar.
The equations of motion following from this are
\ba
\label{eqn:11Dmetriceom}
\hat{\cal R}_{\mu\u}-\frac{1}{12}\left[ F_{\mu\rho_1\rho_2\rho_3}F_\nu^{\;\;\rho_1\rho_2\rho_3}
                                       -\frac{1}{12}g_{\mu\nu}F^2\right]&=&0,\\
\label{eqn:11Dpotentialeom}
d\star F+\half F\wedge F&=&0.
\ea
As we are interested in the case where the internal manifold is a coset, with
squashing parameters that vary as a function of spacetime, we write the
metric as
\ba
\label{eqn:metricAnsatz}
ds^2&=&e^{2\psi(x)}ds^2_{(4)}+g_{ij}(x)e^i\otimes e^j,
\ea
where the one-forms $e^i$ span the cotangent space of the coset manifold
(see appendix \ref{AppCosets}).
We also allow there to be a Freund-Rubin flux \cite{Freund:1980xh} of the form
\ba
\label{eqn:fluxAnsatz}
F&=&f\eta_4,
\ea
where $\eta_4$ is the volume form on spacetime and $f$ is a function to be determined.
In order to have a convenient four dimensional description we require an effective action
in four dimensions which reproduces the equations of motion required to solve the full 11D system. This raises the issue of consistent truncation
which is known to impose certain constraints
on the internal manifold. In the case of group reductions, for example, one
requires the group to be unimodular \cite{Scherk:1979zr}; see \cite{Cvetic:2003jy} for a nice
history and discussion of this issue and see \cite{Pons:2003ka,Pons:2004ky}
for recent work.
For coset reduction we shall find that the effective action for the 
fields from the gravity sector does
actually correspond to simply substituting the 11D metric ansatz into
the 11D action, but the coset must again satisfy a particular constraint.
This issue is not just related to the gravity sector, we also find it in
the flux sector where in order to get the correct sign from the flux contribution
one must appeal to the underlying equations of motion \cite{Duff:1989ah}. 

\section{Choosing a coset}
\label{sec:coset}

The eleven dimensional supergravity permits classical solutions
where the space time is partitioned as $ M_{11} = M_{4} \times M_{7} $
and the internal seven dimensional part is
compact giving an effective theory in four dimensions.   
We are interested in the case where $M_7$ takes the form of a coset manifold,
fortunately such cosets have been classified in
\cite{Castellani:1983yg} and are shown in
table \ref{fig:cosets}. 
We have not explicitly included SO(8)/SO(7) as this is metrically equivalent to
a particular case of the SO(5)/SO(3) coset, where the metric is proportional to
the identity.
Each of these describes
a number of different cosets, depending upon the exact embedding
of the subgroup $H$ in $G$.  For instance the group $SO(5)$ has two
orthogonal $SO(3)$ subgroups, refered to as $A$ and $B$.  We can form a
coset by dividing out by either of these, or by taking some combination
of them.  In this way we find that there are three cosets which are
referred to as ${}_A$, ${}_{A+B}$ and ${}_{MAX}$.  In the same way the M, N and Q
spaces have their subgroup embeddings parametrised by the integers $p,
q$ and $r$.

We now describe the effective theories derived from each type of coset. With all the
effective actions in hand we shall then study their scaling behaviour.

\begin{table}
\center
\begin{tabular}{cccc}
\hline
M${}_7$ & G & H & ref\\
\hline
S${}^7$, J$^7$, V$_{5,2}$ & SO(5) & SO(3) & \cite{Castellani:1983yg}\\
M${}^{pqr}$ & SU(3) $\times$ SU(2) $\times$ U(1) & SU(2) $\times$ U(1)
    $\times$ U(1) & \cite{Castellani:1983mf}\\
N${}^{pqr}$ & SU(3) $\times$ U(1) & U(1) $\times$ U(1) & \cite{Castellani:1983tc} \\
Q${}^{pqr}$ & SU(2) $\times$ SU(2) $\times$ SU(2) & U(1) $\times$ U(1) &
    \cite{D'Auria:1983vy} \\
\hline
\end{tabular}
\caption{
the cosets of 11d supergravity.
}
\label{fig:cosets}
\end{table}

\subsection{Equations for SO(5)/SO(3)$_A$}
\label{sec:SO(3)A}
The general procedure for dimensional reduction on a coset is given in appendix \ref{AppReduceRicci},
here we produce the results for the various allowed cosets presented in table \ref{fig:cosets}.
Our first example is one of the SO(5)/SO(3) cosets, SO(5)/SO(3)$_A$,
presented in appendix \ref{AppSO5SO3A}, and we discover
that there are seven moduli describing the coset metric. However, we need to make sure that the truncation
is consistent, meaning that the 11D Ricci tensor must 
satisfy (\ref{eqn:11Dmetriceom}). If we look at the components of $\hat {\cal R}_{\mu i}$
from (\ref{eqn:mixedRicci}) and use the structure constants relevant for this coset we find that in
general one must
restrict to a diagonal coset metric of the form
\ba
g_{ab}={\rm diag}(e^{2A},e^{2B},e^{2C},e^{2D},e^{2D},e^{2D},e^{2D}).
\ea 
This
is an important restriction, showing that the general metric which respects the coset symmetries does not form
a consistent truncation. This does not however necessarily mean that there are no special configurations which
contain off-diagonal terms for which $\hat {\cal R}_{\mu i}$ happens to vanish.

\begin{figure}
\center
\epsfig{file=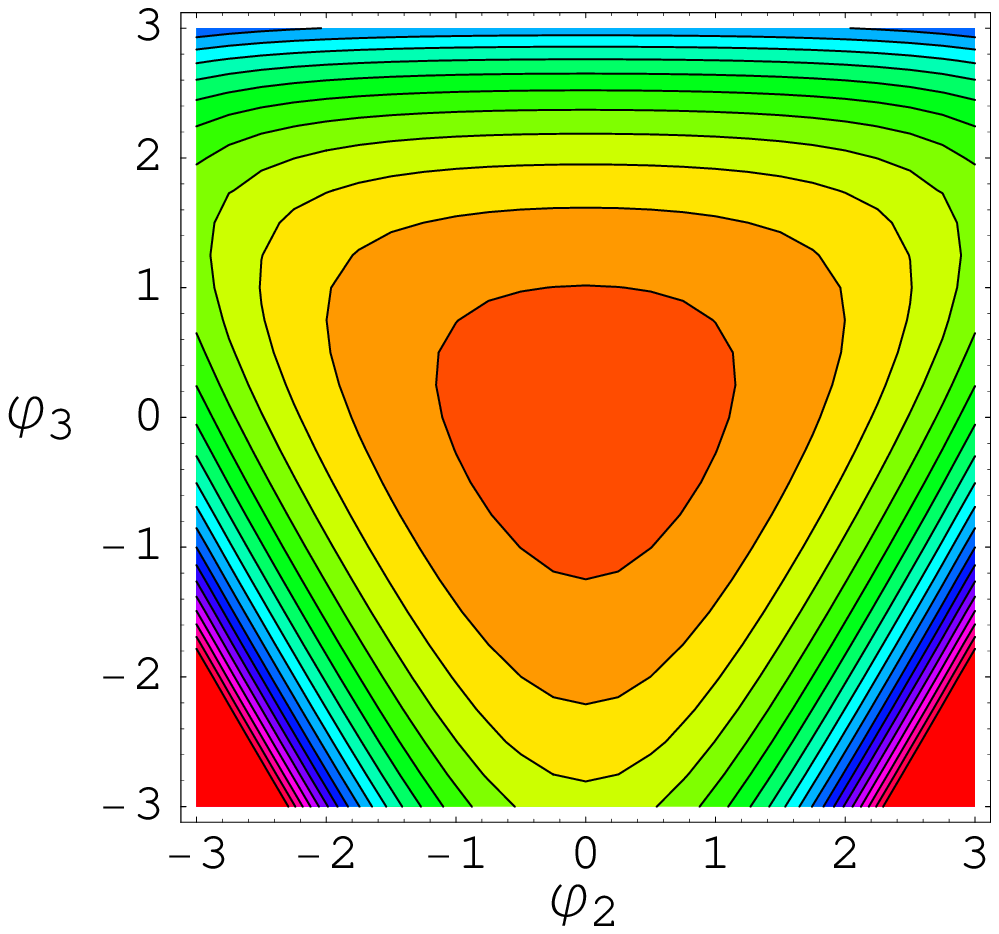,width=6cm}
\epsfig{file=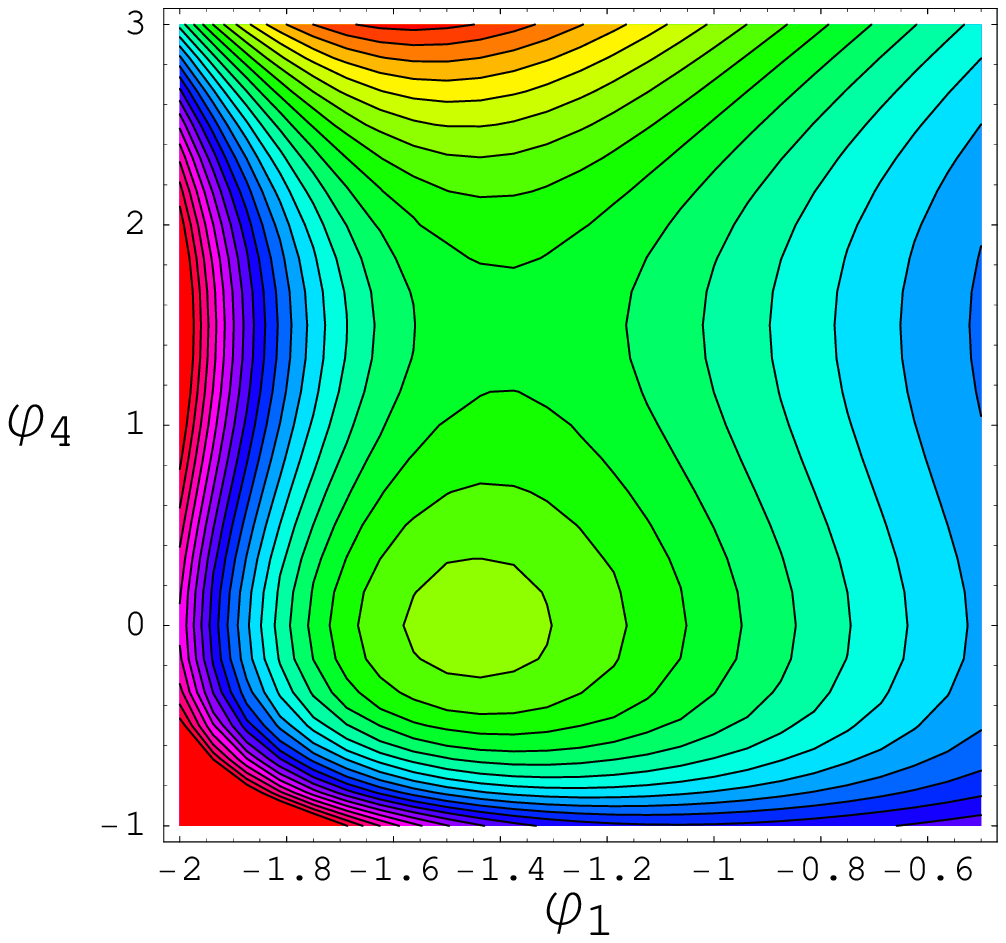,width=6cm}
\flushleft
\caption{
contour plots of the potential for the SO(5)/SO(3)$_A$ coset.
}
\label{fig:SO(5)Apotential}
\end{figure}

For this metric we have from (\ref{eqn:RicciScalarWithGauge}) that
\ba
\label{eqn:Aeofm}
&~&\frac{1}{2\kappa_{11}^2}\int\sqrt{-\hat g_{11}}d^{11}x\hat{\cal R}\\\nonumber
&=&\frac{V_{G/H}}{2\kappa_{11}^2}\int\sqrt{-g_{4}}d^4x\left[ {\cal R}_4+\frac{1}{4}\nabla_\mu g^{ij}\nabla^\mu g_{ij}
                                     -\frac{1}{8}g^{ij}\nabla_\mu g_{ij}g^{kl}\nabla_\mu g_{kl}
                         +e^{2\psi}{\cal R}_{G/H}\right],\\\nonumber
&=&\frac{V_{G/H}}{2\kappa_{11}^2}\int\sqrt{-g_{4}}d^4x
   \Big[ {\cal R}_4-(\nabla A)^2-(\nabla B)^2-(\nabla C)^2-4(\nabla D)^2\\\nonumber
       && \qquad\qquad\qquad\qquad\qquad \qquad\qquad\qquad\qquad -\frac{1}{2}[\nabla(A+B+C+4D)]^2+e^{2\psi}{\cal R}_{G/H}\Big],
\ea
\noindent where we have introduced $V_{G/H}$ which is a {\it constant} volume of the unsquashed coset, defined by
\ba
V_{G/H}&=&\int_{G/H}e^1\wedge e^2\wedge e^3...,
\ea
with the $e^i$ being representatives of the coset cotangent space as in appendix \ref{AppCosets}.
This allows us to define a 4D gravitational coupling for the effective
theory using
\ba
\frac{V_{G/H}}{2\kappa_{11}^2}=\frac{1}{2\kappa^2}.
\ea
We notice from (\ref{eqn:Aeofm}) that the functions $A$, $B$, $C$, $D$ have become scalar
fields of the effective theory, but with non-canonical kinetic terms. In order
to reduce this to standard form we diagonalise the gradient terms in the above expression
using a Gram-Schmidt procedure and introduce
\ba
A&=&\kappa\left(\frac{1}{3}\sqrt{\frac{2}{7}}\varphi_1
               -\frac{1}{\sqrt{2}}\varphi_2
               -\frac{1}{\sqrt{6}}\varphi_3
               -\frac{2}{\sqrt{21}}\varphi_4\right),\\\nonumber
B&=&\kappa\left(\frac{1}{3}\sqrt{\frac{2}{7}}\varphi_1
               +\frac{1}{\sqrt{2}}\varphi_2
               -\frac{1}{\sqrt{6}}\varphi_3
               -\frac{2}{\sqrt{21}}\varphi_4\right),\\\nonumber
C&=&\kappa\left(\frac{1}{3}\sqrt{\frac{2}{7}}\varphi_1
               +\sqrt{\frac{2}{3}}\varphi_3
               -\frac{2}{\sqrt{21}}\varphi_4\right),\\\nonumber
D&=&\kappa\left(\frac{1}{3}\sqrt{\frac{2}{7}}\varphi_1
               +\half\sqrt{\frac{3}{7}}\varphi_4\right).
\ea
Having found the action coming from the Ricci scalar we now consider the dynamics coming from
the Freund-Rubin flux. As we are presently only concerned with flux of the form (\ref{eqn:fluxAnsatz})
the $F\wedge F$ term in (\ref{eqn:11Dpotentialeom}) vanishes to leave
\ba
{\rm d}\left(fe^{-4\psi}e^{A+B+C+4D}\right)&=&0.
\ea
So, using (\ref{eqn:gaugeChoice}), our flux parameter is given by
\ba
\label{eqn:fluxSol}
f=f_0e^{6\psi},
\ea
where $f_0$ is an integration constant.
We are now in a position to examine the equations of motion (\ref{eqn:11Dmetriceom}) finding that they can be
derived from an effective action,
\ba
\label{eqn:potA}
S_4&=&\int\sqrt{-g_4}d^4x\left[ \frac{1}{2\kappa^2}{\cal R}_4
                               -\half\sum_i\nabla_\mu\varphi_i\nabla^\mu\varphi_i
                               -V(\varphi)\right],\\\nonumber
V(\varphi)&=&-\frac{1}{2\kappa^2}e^{-\frac{3}{7}\sqrt{14}\kappa\varphi_1}\left[
               12 e^{-\frac{1}{7}\sqrt{21}\kappa\varphi_4}
              -{\cal V}(-\kappa\varphi_2,-\kappa\varphi_3)e^{-\frac{10}{21}\sqrt{21}\kappa\varphi_4}\right.\\\nonumber
     &~&\hspace{3cm}
        +{\cal V}(\kappa\varphi_2,\kappa\varphi_3)e^{\frac{4}{21}\sqrt{21}\kappa\varphi_4}
              \left.-\half{\cal V}(2\kappa\varphi_2,-2\kappa\varphi_3)e^{\frac{4}{21}\sqrt{21}\kappa\varphi_4}\right]
        +\half f_0^2 e^{-\sqrt{14}\kappa\varphi_1}.
\ea
We have introduced the function
\ba
{\cal V}(x,y)&=&e^{\sqrt{2}x+\sqrt{\frac{2}{3}}y}+e^{-\sqrt{2}x+\sqrt{\frac{2}{3}}y}+e^{-2\sqrt{\frac{2}{3}}y},
\ea
as it aids our understanding of the dynamics in separate
sectors. The function ${\cal V}$ has a global minimum at $\varphi_2=0=\varphi_3$ which
means that
these fields will evolve to zero independently of what $\varphi_1$ and $\varphi_4$ are
doing and we may consider how the potential depends on $\varphi_1$ and $\varphi_4$ alone.
We see from figure \ref{fig:SO(5)Apotential} that there is indeed a minimum at $\varphi_2=\varphi_3=0$.
The figure also shows a local minimum in the $\varphi_1-\varphi_4$ plane along with a saddle point,
these however depend on the value of $\varphi_2$ and $\varphi_3$, with the plot using $\varphi_2=\varphi_3=0$.

We can understand these extrema by studying the existing literature. In the early days of supergravity
there was much interest in existence of Einstein metrics on coset manifolds as these were known to correspond
to extrema of the effective potential. In \cite{Castellani:1983yg} we find that for SO(5)/SO(3)$_A$ there are
two Einstein metrics on the coset, both with $A = B = C$ but with one
taking  $A = D$ and the other with
$A  = D-\log\sqrt{5} $ (\ref{eqn:so3AEinstein}); these correspond to a round seven sphere and a squashed seven sphere respectively
and are given by the local minimum and saddle point of the potential.
In terms of the canonical scalars the round metric takes the values $\varphi_2=\varphi_3=\varphi_4=0$
and the squashed metric has $\varphi_2=\varphi_3=0$, $\varphi_4=\frac{2}{\kappa}\sqrt{\frac{3}{7}}\log\sqrt{5}$. At both the
round sphere and the squashed sphere one finds that the potential has an extremum at which point the
potential is negative, allowing an AdS$_4$ solution.

\subsection{Equations for SO(5)/SO(3)$_{A+B}$}
\label{sec:SO(3)AB}
\begin{figure}
\label{fig:SO(5)A+Bpotential}
\center
\epsfig{file=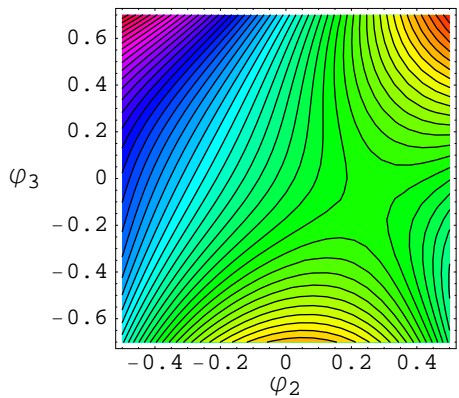,width=6cm}
\flushleft
\caption{
contour plots of the potential for the SO(5)/SO(3)$_{A+B}$ coset.
}
\label{fig:SO(5)ABpotential}
\end{figure}

As mentioned in section \ref{sec:SO(3)A} there is another coset of SO(5) where the SO(3) subgroup is
taken to be the diagonal component of SO(3)$_A\otimes$SO(3)$_B$. In appendix \ref{AppSO5SO3A+B}
we show how such a coset is constructed and derive the general metric which is consistent with the coset
symmetries.
We restrict to a diagonal coset metric $g_{ab}={\rm diag}(e^{2A},e^{2A},e^{2A},e^{2B},e^{2C},e^{2C},e^{2C})$,
which is needed for a consistent reduction.
Following the same procedure as before we calculate from (\ref{eqn:RicciScalarWithGauge}) the effective
action coming from the Ricci scalar,
\ba
&~&\frac{1}{2\kappa_{11}^2}\int\sqrt{-\hat g_{11}}d^{11}x\hat{\cal R}\\\nonumber
&=&\frac{V_{G/H}}{2\kappa_{11}^2}\int\sqrt{-g_{4}}d^4x
   \left[ {\cal R}_4-3(\nabla A)^2-(\nabla B)^2-3(\nabla C)^2
       -\frac{1}{2}[\nabla(3A+B+3C)]^2+e^{2\psi}{\cal R}_{G/H}\right].
\ea
Again we see that the parameters describing the size and shape of the internal manifold have become
scalar fields with non-canonical kinetic terms, and in order
to have standard kinetic terms we diagonalise the gradient terms as we did before. We find that the Gram-Schmidt
procedure provides us with the following field redefinition,
\ba
A&=&\kappa\left(\frac{1}{3}\sqrt{\frac{2}{7}}\varphi_1
               -\sqrt{\frac{1}{12}}\varphi_2
               -\frac{1}{2}\sqrt{\frac{3}{7}}\varphi_3\right),\\\nonumber
B&=&\kappa\left(\frac{1}{3}\sqrt{\frac{2}{7}}\varphi_1
               +\frac{\sqrt{3}}{2}\varphi_2
               -\frac{1}{2}\sqrt{\frac{3}{7}}\varphi_3\right),\\\nonumber
C&=&\kappa\left(\frac{1}{3}\sqrt{\frac{2}{7}}\varphi_1
               +\frac{2}{\sqrt{21}}\varphi_3\right).
\ea
Turning our attention to the equations for the three-form potential
(\ref{eqn:11Dpotentialeom}), we find that the Freund-Rubin flux must satisfy
\ba
{\rm d}\left(fe^{-4\psi}e^{3A+B+3C}\right)&=&0,
\ea
and so using (\ref{eqn:gaugeChoice}) our flux parameter is given by the same expression as before, (\ref{eqn:fluxSol}).
From this one may find the equations of motion and we discover that they can be
derived from the following effective action,
\ba
S_4&=&\int\sqrt{-g_4}d^4x\left[ \frac{1}{2\kappa^2}{\cal R}_4
                               -\half\sum_i\nabla_\mu\varphi_i\nabla^\mu\varphi_i
                               -V(\varphi)\right],\\
\label{eqn:so3A+Bpot}
V(\varphi)&=&-\frac{1}{2\kappa^2}e^{-3\sqrt{\frac{2}{7}}\kappa\varphi_1}\Big[
               9 e^{-\frac{4}{\sqrt{21}}\kappa\varphi_3}
              +9 e^{\frac{1}{\sqrt{3}}\kappa\varphi_2
		    + \sqrt{\frac{3}{7}}\kappa\varphi_3}
              +3 e^{-\sqrt{3}\kappa\varphi_2
		    + \sqrt{\frac{3}{7}}\kappa\varphi_3}\\\nonumber
	&~&
              -\frac{3}{2} e^{-\frac{4}{\sqrt{3}}\kappa\varphi_2
		    -\frac{4}{\sqrt{21}}\kappa\varphi_3}
              -\frac{3}{2} e^{\frac{4}{\sqrt{3}}\kappa\varphi_2
		    -\frac{4}{\sqrt{21}}\kappa\varphi_3}
              -\frac{3}{2} e^{-\frac{2}{\sqrt{3}}\kappa\varphi_2
		    +\frac{10}{\sqrt{21}}\kappa\varphi_3}
	\Big]
        +\half f_0^2 e^{-\sqrt{14}\kappa\varphi_1}.
\ea
On examining the contour plot of this potential in the
$\varphi_2-\varphi_3$ plane (figure \ref{fig:SO(5)A+Bpotential})  we observe
that there is an extremum, as is to be expected, which corresponds to the location where the metric is
Einstein \cite{Castellani:1983yg};
this point is independent of the scalar $\varphi_1$. The Einstein metric is derived in appendix \ref{AppSO5SO3A+B}
and is given by
\ba
    \varphi_2 &=& \frac{\sqrt{3}}{4\kappa}  \log(\frac{3}{2}),\nonumber\\
    \varphi_3 &=& - \frac{1}{\sqrt{7}} \varphi_2,
\ea
at which point the potential becomes
\ba
V(\varphi)&=&-\frac{1}{2\kappa^2}\frac{63}{2}\left(\frac{3}{2}\right)^{1/7}e^{-3\sqrt{\frac{2}{7}}\kappa\varphi_1}
+\half f_0^2 e^{-\sqrt{14}\kappa\varphi_1},
\ea
which allows for an AdS$_4$ solution in the minimum.

\subsection{Equations for SO(5)/SO(3)$_{MAX}$}
\label{sec:SO(3)MAX}
The last coset one can construct out of SO(3) subgroups of SO(5) is where the SO(3) subgroup is maximal. We construct
this subgroup in appendix \ref{AppSO5SO3MAX} finding only a
single modulus, which describes just a breathing mode
for the internal space and therefore there are no shape moduli.
We are only allowed the single parameter diagonal coset metric which we parametrise as
$g_{ab}={\rm diag}(e^{2A},e^{2A},e^{2A},e^{2A},e^{2A},e^{2A},e^{2A})$
and for which we calculate from (\ref{eqn:RicciScalarWithGauge})
\ba
\frac{1}{2\kappa_{11}^2}\int\sqrt{-\hat g_{11}}d^{11}x\hat{\cal R}
&=&\frac{V_{G/H}}{2\kappa_{11}^2}\int\sqrt{-g_{4}}d^4x
   \left[ {\cal R}_4-\frac{63}{2}(\nabla A)^2
	+e^{2\psi}{\cal R}_{G/H}\right].
\ea
These are the terms which will give us the kinetic terms for the effective action. In order
to have canonical kinetic terms we rescale $A$ as follows,
\be
A =\frac{1}{3}\sqrt{\frac{2}{7}}\kappa \varphi_1.
\ee
The effect of the Freund-Rubin flux is calculated as before, with
(\ref{eqn:11Dpotentialeom}) giving
\ba
{\rm d}\left(fe^{-4\psi}e^{7A}\right)&=&0,
\ea
and once again our flux parameter is given by (\ref{eqn:fluxSol}).
We find that the evolution of the moduli coming from 11D equations of motion 
can then be described by the following effective action,
\ba
S_4&=&\int\sqrt{-g_4}d^4x\left[ \frac{1}{2\kappa^2}{\cal R}_4
                               -\half\sum_i\nabla_\mu\varphi_i\nabla^\mu\varphi_i
                               -V(\varphi)\right],\\\nonumber
V(\varphi)&=&-\frac{1}{2\kappa^2}\frac{189}{10}e^{-3\sqrt{\frac{2}{7}}\kappa\varphi_1}
        +\half f_0^2 e^{-\sqrt{14}\kappa\varphi_1},
\ea

\noindent which has an AdS$_4$ extremum at $\varphi_1 = \frac{1}{4\kappa} \sqrt{\frac{7}{2}}
\log(\frac{10 {f_0}^2 \kappa^2}{81})$.

\begin{figure}
\center
\epsfig{file=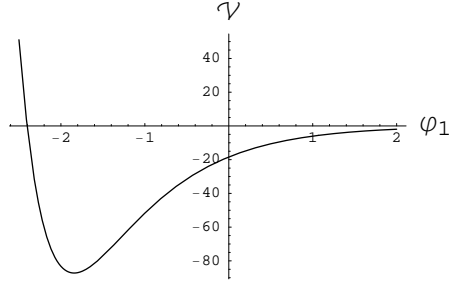,width=6cm}
\flushleft
\caption{
Plot of the potential for the SO(5)/SO(3)$_{MAX}$ coset.
}
\label{fig:SO(5)MAXpotential}
\end{figure}

\subsection{Equations for M${}^{pqr}$ = SU(3) x SU(2) x U(1)/SU(2) x U(1) x U(1)}
\label{sec:Mpqr}
We now consider another class of cosets, SU(3) x SU(2) x U(1) /
SU(2) x U(1) x U(1), characterised by integers $p, q, r$, the
construction of which we show in appendix \ref{AppMpqr}.
A full discussion of these spaces can be found in
\cite{Castellani:1983yg,Witten:1981me,Page:1984ad}, in particular
the curvature depends only on the ratio $p/q$, and the space M$^{pq0}$ is the
covering space of M$^{pqr}$.

We parametrise the metric for this class of coset (\ref{Mmetric}) in
maximum generality as the diagonal metric $g_{ab}={\rm
diag}(e^{2A},e^{2A},e^{2A},e^{2A},e^{2B},e^{2B},e^{2C})$, and following 
the same procedure as before we calculate from (\ref{eqn:RicciScalarWithGauge}) 
the effective action coming from the Ricci scalar,
\ba
&~&\frac{1}{2\, \kappa_{11}^2}\int\, \sqrt{-\hat g_{11}}\, d^{11}x\, \hat{\cal R}\\\nonumber
&=&\frac{V_{G/H}}{2\, \kappa_{11}^2}\int\, \sqrt{-g_{4}}\, d^4x \,
   \left[ {\cal R}_4-4(\nabla A)^2-2(\nabla B)^2-(\nabla C)^2
	-2[\nabla(2A+B+\frac{1}{2}\,C)]^2+e^{2\,\psi}{\cal
R}_{G/H}\right].
\ea
Again we see that the parameters describing the size and shape of the internal manifold have become
scalar fields with non-canonical kinetic terms, and so in order
to have standard kinetic terms we diagonalise the gradient terms as we did before. We find that the Gram-Schmidt
procedure provides us with the following field redefinition,
\ba
A&=&\kappa\left(\frac{1}{3}\, \sqrt{\frac{2}{7}}\, \varphi_1
		-\frac{1}{\sqrt{6}}\, \varphi_2
		-\frac{1}{\sqrt{42}}\, \varphi_3\right)
		+\frac{1}{9} \log(\frac{q}{\zeta}),\\\nonumber
B&=&\kappa\left(\frac{1}{3}\, \sqrt{\frac{2}{7}}\, \varphi_1
		+\frac{1}{\sqrt{3}}\, \varphi_2
		-\frac{1}{\sqrt{42}}\, \varphi_3\right)
		+\frac{1}{9} \log(\frac{q}{\zeta}),\\\nonumber
C&=&\kappa\left(\frac{1}{3}\, \sqrt{\frac{2}{7}}\, \varphi_1
		+\sqrt{\frac{6}{7}}\, \varphi_3\right)
		-\frac{8}{9} \log(\frac{q}{\zeta}),
\ea
\noindent with $\zeta = \sqrt{3\, p^2 + q^2 + 2 r^2}$. In these expressions we have also shifted
the fields by a constant factor to simplify the potential.

We find that the Freund-Rubin flux must satisfy
\ba
{\rm d}\left(fe^{-4\, \psi}e^{4\, A+2\, B+C}\right)&=&0,
\ea
and so using (\ref{eqn:gaugeChoice}), our flux parameter is given by the same expression as before, (\ref{eqn:fluxSol}).
From this one may find the equations of motion and we discover that they can be
derived from the following effective action,
\ba
S_4&=&\int\sqrt{-g_4}\, d^4x\, \left[ \frac{1}{2\, \kappa^2}\,{\cal R}_4
                               -\half\sum_i\nabla_\mu\varphi_i\, \nabla^\mu\varphi_i
                               -V(\varphi)\right],\\
\label{eqn:Mpqr}
V(\varphi)&=&-\frac{1}{2\, \kappa^2}e^{-3\, \sqrt{\frac{2}{7}}\kappa\,\varphi_1}\Big[
	      -\frac{9}{4}\frac{p^2}{q^2} e^{\frac{2}{\sqrt{3}}\kappa\,\varphi_2
		    + 8\,\sqrt{\frac{2}{21}}\kappa\,\varphi_3}
              +2\, e^{-\frac{2}{\sqrt{3}}\kappa\,\varphi_2
		    + \sqrt{\frac{2}{21}}\kappa\,\varphi_3}\\\nonumber
	&~&
              +6\, e^{\frac{1}{\sqrt{3}}\kappa\,\varphi_2
		    + \sqrt{\frac{2}{21}}\kappa\,\varphi_3}
              -\half e^{-\frac{4}{\sqrt{3}}\kappa\,\varphi_2
		    +8\,\sqrt{\frac{2}{21}}\kappa\,\varphi_3}
	\Big]
        +\half \left(\frac{q^2}{3\, p^2 + q^2 + 2\, r^2}\right)^{1/3}
	    \, f_0^2 \, e^{-\sqrt{14}\kappa\,\varphi_1}.
\ea

\subsection{Equations for N${}^{pqr}$ = SU(3) x U(1)/U(1) x U(1)}
\label{sec:Npqr}
Another class of cosets to consider are SU(3) x U(1) / U(1) x U(1), 
which are also characterised by integers $p, q, r$;
the construction is shown in appendix \ref{AppNpqr}, but further details may be found in
\cite{Page:1984ac,Castellani:1983tc}. For this coset one finds again that
the curvature is independent of $r$, with N${}^{pq0}$ being the covering 
space of N${}^{pqr}$.

The general metric for this class of coset is again diagonal
(\ref{Nmetric}), which by taking
$g_{ab}={\rm diag}(e^{2A},e^{2A},e^{2B},e^{2B},e^{2C},e^{2C},e^{2D})$
and calculating as before from (\ref{eqn:RicciScalarWithGauge}) we obtain the 
effective action coming from the Ricci scalar,
\ba
&~&\frac{1}{2\kappa_{11}^2}\int\sqrt{-\hat g_{11}}d^{11}x\hat{\cal R}\\\nonumber
&=&\frac{V_{G/H}}{2\kappa_{11}^2}\int\sqrt{-g_{4}}d^4x
   \left[ {\cal R}_4-2(\nabla A)^2-2(\nabla B)^2-2(\nabla C)^2
       -2[\nabla(A+B+C+\frac{1}{2}D)]^2+e^{2\psi}{\cal R}_{G/H}\right].
\ea
Introducing a constant shift in the field redefinition following from the Gram-Schmidt
process we find the following gives canonical kinetic terms for the scalars $\varphi_i$,
\ba
A&=&\kappa\left(\frac{1}{3}\sqrt{\frac{2}{7}}\varphi_1
		-\frac{1}{2}\varphi_2
		-\frac{1}{\sqrt{6}}\varphi_3
		-\frac{1}{\sqrt{42}}\varphi_4\right)
		+\frac{1}{9} \log(\frac{q}{\zeta}),\\\nonumber
B&=&\kappa\left(\frac{1}{3}\sqrt{\frac{2}{7}}\varphi_1
		+\frac{1}{2}\varphi_2
		-\frac{1}{\sqrt{6}}\varphi_3
		-\frac{1}{\sqrt{42}}\varphi_4\right)
		+\frac{1}{9} \log(\frac{q}{\zeta}),\\\nonumber
C&=&\kappa\left(\frac{1}{3}\sqrt{\frac{2}{7}}\varphi_1
		+\frac{1}{\sqrt{3}}\varphi_3
		-\frac{1}{\sqrt{42}}\varphi_4\right)
		+\frac{1}{9} \log(\frac{q}{\zeta}),\\\nonumber
D&=&\kappa\left(\frac{1}{3}\sqrt{\frac{2}{7}}\varphi_1
		+\sqrt{\frac{6}{7}}\varphi_4\right)
		-\frac{8}{9} \log(\frac{q}{\zeta}),
\ea
\noindent with $\zeta = \sqrt{3\, p^2 + q^2 + 2 r^2}$.

The flux equations of motion require the Freund-Rubin flux to satisfy
\ba
{\rm d}\left(fe^{-4\psi}e^{2A+2B+2C+D}\right)&=&0,
\ea
and we discover that the equations of motion can be
derived from the effective action,
\ba
S_4&=&\int\sqrt{-g_4}d^4x\left[ \frac{1}{2\kappa^2}{\cal R}_4
                               -\half\sum_i\nabla_\mu\varphi_i\nabla^\mu\varphi_i
                               -V(\varphi)\right]\\
\label{eqn:Npqr}
V(\varphi)&=&-\frac{1}{2\kappa^2}e^{-3\sqrt{\frac{2}{7}}\kappa\varphi_1}\Big[
	      -\frac{(3p-q)^2}{8 q^2} e^{-\frac{4}{\sqrt{3}}\kappa\varphi_3
		    + 8\sqrt{\frac{2}{21}}\kappa\varphi_4}
              +3 e^{-\frac{2}{\sqrt{3}}\kappa\varphi_3
		    + \sqrt{\frac{2}{21}}\kappa\varphi_4}\\\nonumber
	&~&
              -\frac{1}{2} e^{-2\kappa\varphi_2
		    - \frac{2}{\sqrt{3}}\kappa\varphi_3
		    + \sqrt{\frac{2}{21}}\kappa\varphi_4}
              -\frac{1}{2} e^{2\kappa\varphi_2
		    -\frac{2}{\sqrt{3}}\kappa\varphi_3
		    +\sqrt{\frac{2}{21}}\kappa\varphi_4}
              + 3 e^{-\kappa\varphi_2
		    +\frac{1}{\sqrt{3}}\kappa\varphi_3
		    +\sqrt{\frac{2}{21}}\kappa\varphi_4}\\\nonumber
	&~&
              + 3 e^{\kappa\varphi_2
		    +\frac{1}{\sqrt{3}}\kappa\varphi_3
		    +\sqrt{\frac{2}{21}}\kappa\varphi_4}
              - \frac{1}{2} e^{\frac{4}{\sqrt{3}}\kappa\varphi_3
		    + \sqrt{\frac{2}{21}}\kappa\varphi_4}
              - \frac{1}{2} e^{2\kappa\varphi_2
		    +\frac{2}{\sqrt{3}}\kappa\varphi_3
		    +8\sqrt{\frac{2}{21}}\kappa\varphi_4}\\\nonumber
	&~&
	      -\frac{(3p+q)^2}{8 q^2} e^{-2\kappa\varphi_2
		    + \frac{2}{\sqrt{3}}\kappa\varphi_3
		    + 8\sqrt{\frac{2}{21}}\kappa\varphi_4}
	\Big]
        +\half \left(\frac{q^2}{3\, p^2 + q^2 + 2\, r^2}\right)^{1/3}\,
	    f_0^2 \, e^{-\sqrt{14}\kappa\varphi_1}.
\ea

\subsection{Equations for Q${}^{pqr}$ = SU(2) x SU(2) x SU(2)/U(l) x U(1)}
\label{sec:Qpqr}
The final class of cosets in the classification are the 
Q${}^{pqr}$ spaces, given by the quotient SU(2)
x SU(2) x SU(2) / U(1) x U(1) where the integers $p, q, r$ characterise the
different embeddings of the subgroup. This coset is constructed in
appendix \ref{AppQpqr},
but further details on its structure may be found in \cite{Page:1984ae}.

The metric for this class of coset is the diagonal metric
$g_{ab}={\rm diag}(e^{2A},e^{2A},e^{2B},e^{2B},e^{2C},e^{2C},e^{2D})$
for which we calculate from
(\ref{eqn:RicciScalarWithGauge}) that the effective action coming from
the Ricci scalar is,
\ba
&~&\frac{1}{2\kappa_{11}^2}\int\sqrt{-\hat g_{11}}d^{11}x\hat{\cal R}\\\nonumber
&=&\frac{V_{G/H}}{2\kappa_{11}^2}\int\sqrt{-g_{4}}d^4x
   \Big[ {\cal R}_4-2(\nabla A)^2-2(\nabla B)^2-(\nabla C)^2-(\nabla D)^2\\\nonumber
    && \qquad\qquad\qquad\qquad\qquad \qquad\qquad\qquad\qquad
       -2[\nabla(A+B+C+\frac{1}{2}D)]^2+e^{2\psi}{\cal R}_{G/H}\Big].
\ea
To get back to canonical kinetic terms we introduce field redefinitions,
\ba
A&=&\kappa\left(\frac{1}{3}\sqrt{\frac{2}{7}}\varphi_1
		-\frac{1}{2}\varphi_2
		-\frac{1}{\sqrt{6}}\varphi_3
		-\frac{1}{\sqrt{42}}\varphi_4\right)
		+\frac{1}{9} \log(\frac{q}{\zeta}),\\\nonumber
B&=&\kappa\left(\frac{1}{3}\sqrt{\frac{2}{7}}\varphi_1
		+\frac{1}{2}\varphi_2
		-\frac{1}{\sqrt{6}}\varphi_3
		-\frac{1}{\sqrt{42}}\varphi_4\right)
		+\frac{1}{9} \log(\frac{q}{\zeta}),\\\nonumber
C&=&\kappa\left(\frac{1}{3}\sqrt{\frac{2}{7}}\varphi_1
		+\frac{1}{\sqrt{3}}\varphi_3
		-\frac{1}{\sqrt{42}}\varphi_4\right)
		+\frac{1}{9} \log(\frac{q}{\zeta}),\\\nonumber
D&=&\kappa\left(\frac{1}{3}\sqrt{\frac{2}{7}}\varphi_1
		+\sqrt{\frac{6}{7}}\varphi_4\right)
		-\frac{8}{9} \log(\frac{q}{\zeta}),
\ea
\noindent with $\zeta = \sqrt{p^2 + q^2 + r^2}$,
and the flux equation gives us
\ba
{\rm d}\left(fe^{-4\psi}e^{2A+2B+2C+D}\right)&=&0.
\ea
From this one may find the equations of motion and we discover that they can be
derived from the following effective action,
\ba
S_4&=&\int\sqrt{-g_4}d^4x\left[ \frac{1}{2\kappa^2}{\cal R}_4
                               -\half\sum_i\nabla_\mu\varphi_i\nabla^\mu\varphi_i
                               -V(\varphi)\right]\\
\label{eqn:Qpqr}
V(\varphi)&=&-\frac{1}{2\kappa^2}e^{-3\sqrt{\frac{2}{7}}\kappa\varphi_1}\Big[
	      -\frac{p^2}{2 q^2} e^{2\kappa\varphi_2
		    + \frac{2}{\sqrt{3}}\kappa\varphi_3
		    + 8\sqrt{\frac{2}{21}}\kappa\varphi_4}
              + 2 e^{-\frac{2}{\sqrt{3}}\kappa\varphi_3
		    + \sqrt{\frac{2}{21}}\kappa\varphi_4}\\\nonumber
	&~&
              + 2 e^{-\kappa\varphi_2
		    + \frac{1}{\sqrt{3}}\kappa\varphi_3
		    + \sqrt{\frac{2}{21}}\kappa\varphi_4}
              + 2 e^{\kappa\varphi_2
		    +\frac{1}{\sqrt{3}}\kappa\varphi_3
		    +\sqrt{\frac{2}{21}}\kappa\varphi_4}
              - \half e^{-2\kappa\varphi_2
		    +\frac{2}{\sqrt{3}}\kappa\varphi_3
		    +8\sqrt{\frac{2}{21}}\kappa\varphi_4}\\\nonumber
	&~&
	      -\frac{r^2}{2 q^2} e^{
		    - \frac{4}{\sqrt{3}}\kappa\varphi_3
		    + 8\sqrt{\frac{2}{21}}\kappa\varphi_4}
	\Big]
        +\half \left(\frac{q^2}{p^2+q^2+r^2}\right)^{1/3}
	    \, f_0^2 \, e^{-\sqrt{14}\kappa\varphi_1}.
\ea

\section{Scaling solutions}
\label{sec:scaling}
Now that we have our set of effective actions for the full set of cosets we
move on to examine their dynamics.
While studying the different possible types of evolution a scalar field may have it
becomes clear that scaling solutions coming from exponential potentials
hold a rather prominent position
\cite{Copeland:1997et,Collinucci:2004iw,Wetterich:1987fm,Burd:1988ss,Kaloper:1991mq,Brustein:1992nk,Barreiro:1998aj,Barreiro:2000pf,Brustein:2004jp,Barreiro:2005ua,Coley:1999mj,Blais:2004vt}. This is because systems of exponential potentials
can have attractor solutions, when viewed from a dynamical systems perspective \cite{Copeland:1997et,Collinucci:2004iw},
where the scalar field constitutes a fixed fraction of the energy density in the
Universe \cite{Kim:2005ne}. We have seen in the above sections that coset spaces naturally give rise to
exponential potentials, a situation which is common in dimensionally reduced theories. Another nice
property of having many scalars is that while each term in the potential may be too steep to support
an accelerating scale factor, together they may combine in such a way as to allow inflationary
behaviour \cite{Liddle:1998jc}. We can understand this by noting that in some cases it is possible
to perform a field redefinition such that only one of the new fields is involved in the evolution,
or some smaller subset than was being used in the initial formulation \cite{Collinucci:2004iw}.
In this section we shall investigate how the scalars evolve for the coset reductions discussed above,
we start with a general discussion of how scaling solutions appear. In doing this we shall follow
\cite{Barreiro:1998aj,Barreiro:2005ua} by adding a fluid component which accounts for
other forms of matter/radiation that could be present.
We describe this fluid by the following equation of state,
\ba
P_\gamma&=&(\gamma-1)\rho_\gamma.
\ea
Taking the cosmological ansatz of homogeneous fields, $\phi(x^\mu)=\phi(t)$ and the
FRW metric
\ba
{\rm d}s^2&=&-{\rm d}t^2+a(t)^2\underline{{\rm d}x}^2,
\ea
for flat spatial sections we find the usual equations
\ba
H^2&=&\frac{\kappa^2}{3}\left[\sum_i\half\dot\varphi^2_i+V+\rho_\gamma\right],\\
\dot H&=&-\frac{\kappa^2}{2}\left[\sum_i\dot\varphi^2_i+\gamma\rho_\gamma\right],\\
\dot\rho_\gamma&=&-3\gamma H\rho_\gamma\\
\ddot\varphi_i+3H\dot\varphi+V_{,i}&=&0.
\ea
A central part of scaling solutions is that all terms in a given equation have the
same behaviour. Given that, it is easy to see the above equations require
\ba
H^2\propto\dot\varphi_i^2\propto V\propto\rho_\gamma\propto\dot H,
\ea
which can be consistent only for exponential potentials, and $H\propto 1/t$.
A crucial point in searching for these scaling solutions is that one cannot have
more independent terms in the potential than scalar fields, otherwise the system
is overdetermined and has no scaling solution. We see straight away that for our system we
always have more terms than scalars and so there cannot be any true scaling solution.
That does not rule out there being regions where the evolution is well approximated
by scaling solutions, and as pointed out in \cite{Barreiro:2005ua} this can in fact be
a useful tool, with the evolution moving from scaling solution to scaling solution.
Given that there is no true scaling solution we shall now consider regions of parameter
space where we expect approximate scaling behaviour.

As an example, consider the case of a scalar field with potential 
\ba
\label{eqn:scalingPot}
{\cal V}=A\exp(-b\kappa\phi),
\ea
$A>0$, scale factor with $a(t)\sim t^p$ and fluid density $\rho=\tilde\rho/(\kappa^2 t^2)$.
We find that
\ba
p=2/(3\gamma),\quad \tilde\rho=3p(p-2/b^2)
\ea
with a fluid and
\ba
p=2/b^2
\ea
without the fluid. There are also the consistency relations
\ba
\label{eqn:fluidCon}
b^2>3\gamma
\ea
with a fluid and
\ba
\label{eqn:scalarCon}
b^2<6
\ea
without. These come from demanding positivity of the fluid energy density and potential ${\cal V}$.
If we were to allow $A<0$ then the scaling solution, without fluid, would have the opposite bound
for $b$, $b^2>6$. However, in this case the solution corresponds to the scalar field rolling
up the potential due to a large initial velocity, and is an unstable evolution. Because of this we shall
only be trying to construct scaling solutions from positive terms in the potential.

\section{Scaling regions}
\label{sec:regions}
Taking a general potential of the form\cite{Collinucci:2004iw},
\ba
V&=& \sum_{i=1}^{N} V_i\\
\label{eqn:alphaVectors}
V_i&=& \Lambda_i\exp(\underline\alpha_i.\underline\varphi)
    \,=	\, \Lambda_i\exp(\sum_{I=1}^{n} \alpha_{iI} \varphi_I),
\ea

\noindent
with $n$ scalar fields and $N$ terms then scaling behaviour can be expected only for
potentials with $N\leq n$, otherwise we find an over-determined system
of equations as described in the last section. In our system we have more terms than fields,
therefore we look for directions in field space which can support scaling solutions
owing to a subset of terms being subdominant.
For term $V_{d}$ to dominate over term $V_{s}$ in direction $\underline\Phi$ we need
\ba
\label{eqn:condition1}
\underline\alpha_{s}.\underline\Phi<\underline\alpha_{d}.\underline\Phi.
\ea
In order for scaling to be achieved without going to very large field values we shall require
the subdominant terms ($V_{s}$) to die off exponentially, while the dominant terms ($V_{d}$) increase
exponentially
\ba
\label{eqn:condition2}
\underline\alpha_{d}.\underline\Phi>0,\qquad \underline\alpha_{s}.\underline\Phi<0.
\ea
Furthermore, we guarantee that the scaling solution rolls down the potential,
rather than up, by enforcing the
terms with negative $\Lambda_{i}$ to be subdominant terms. We call all these requirements the
conditions for {\it strong scaling}.

\subsection{SO(5)/SO(3)$_A$}
In order to make the procedure clearer we shall start with a special, restricted, example by
choosing the coset SO(5)/SO(3)$_A$, taking the restriction $\varphi_2=0=\varphi_3$
(\ref{eqn:potA}). Then we find the terms in the potential are given
by the following $\underline\alpha$ vectors (\ref{eqn:potA})(\ref{eqn:alphaVectors}), with 
those contributing to negative terms ($\Lambda_i<0$)
indicated with $\bullet$ marks,
\ba
\bullet\; \underline\alpha_1&=&\frac{\sqrt 7}{21}\kappa(-9\sqrt 2,0,0,-3\sqrt 3),\\\nonumber
\underline\alpha_2&=&\frac{\sqrt 7}{21}\kappa(-9\sqrt 2,0,0,-10\sqrt 3),\\\nonumber
\bullet\; \underline\alpha_3&=&\frac{\sqrt 7}{21}\kappa(-9\sqrt 2,0,0,4\sqrt 3),\\\nonumber
\underline\alpha_4&=&\frac{\sqrt 7}{21}\kappa(-21\sqrt 2,0,0,0).
\ea
The procedure now is to find a direction in field space, $\underline\Phi$, which satisfies
\ba
\underline\alpha_1.\underline\Phi=-\eta^2,\\
\underline\alpha_3.\underline\Phi=-\zeta^2,
\ea
where $\eta$ and $\zeta$ are real constants, thereby ensuring that the
negative terms, $V_1$ and $V_3$,
are exponentially small in the direction $\underline\Phi$. With these conditions we now
parametrise $\underline\Phi$ as
\ba
\Phi&=&\frac{21}{\sqrt{7}\kappa}(\sqrt{2}a,0,0,\sqrt{3}b),
\ea
where the numerical factors are simply for convenience, and we then derive that
\ba
\underline\alpha_2.\underline\Phi&=&-2\eta^2+\zeta^2,\\
\underline\alpha_4.\underline\Phi&=&-\frac{4}{3}\eta^2-\zeta^2.
\ea
We therefore conclude that if the negative terms in the potential, $V_1$
and $V_3$, are exponentially
small then so is $V_4$ as $\underline\alpha_4.\underline\Phi<0$. However, there are regions where
$V_2$ can be large as $\underline\alpha_2.\underline\Phi$ will be positive if $\zeta^2>2\eta^2$.
In this example we find that if
\ba
\label{scaling:bounds:A}
a+2b>0,\qquad 3a+5b<0
\ea
then the term $V_2$ will be exponentially large with $V_1$, $V_3$, $V_4$ being exponentially
small. Thus, this system meets the requirements of {\it strong scaling} as given earlier.

As an example we consider an initial condition
commensurate with (\ref{scaling:bounds:A}) by chosing as a direction for
$\Phi$,
\ba
\Phi=\frac{1}{\sqrt{5}}(\sqrt{2}\varphi_1-\sqrt{3}\varphi_4),\\\nonumber
\chi=\frac{1}{\sqrt{5}}(\sqrt{3}\varphi_1+\sqrt{2}\varphi_4),
\ea
where we have introduced the orthogonal partner, $\chi$, to $\Phi$. 
Note that as this is an
SO(2) rotation the kinetic terms remain canonical.
Substituting this into the potential (\ref{eqn:potA}) we find that the large $\Phi$
limit is described by
\ba
V\sim\frac{3}{2\kappa^2}\exp(\frac{4\sqrt{7}}{7\sqrt{5}} \kappa\Phi-\frac{19\sqrt{42}}{21\sqrt 5}\kappa\chi).
\ea
Performing one more SO(2) field redefinition,
\ba
\psi_1=\sqrt{3/22}\left(\frac{4\sqrt{7}}{7\sqrt{5}} \Phi-\frac{19\sqrt{42}}{21\sqrt 5}\chi\right),\quad 
\psi_2=\sqrt{3/22}\left(\frac{4\sqrt{7}}{7\sqrt{5}} \chi+\frac{19\sqrt{42}}{21\sqrt 5}\Phi\right)
\ea
the potential becomes
\ba
\label{eqn:so3Alimit}
V\sim\frac{3}{2\kappa^2}\exp(\sqrt{22/3} \kappa\psi_1).
\ea
We now see that (\ref{eqn:scalingPot})(\ref{eqn:fluidCon})(\ref{eqn:scalarCon})
imply that a scaling solution is possible, but only in the presence of a
background fluid.  Note that in this example whenever $V_2$ dominates
over the other terms
it is always possible to perform a field
redefinition so that the effective potential will take exactly the same
form as (\ref{eqn:so3Alimit}), showing that tracking solutions are allowed in the region
defined by (\ref{scaling:bounds:A}).

If we now consider the more general case with $\varphi_2$ and $\varphi_3$ non-vanishing
then a similar analysis shows that there can be up to four positive terms dominant in
the potential, as there are four scalars this allows there to be scaling solutions
and the full system can meet the requirements of {\it strong scaling}. 
The terms which can dominate in these more general regions are given in 
table \ref{fig:scalingTerms}. This table lists the terms which are dominant in some
particular directions, for example: there are directions in the SO(5)/SO(3)$_A$ case where
only the $V_2$ increases exponentially; there are also directions where both $V_2$ and $V_3$
increase with the rest decreasing; there are, however, no directions where $V_2$ and $V_8$
increase with the rest decreasing.
We shall defer the study of these
more complicated directions for future work.

\subsection{SO(5)/SO(3)$_{A+B}$}
We can perform a similar analysis for the potential (\ref{eqn:so3A+Bpot}),
for which there are three scalar fields so we are allowed at most three
dominant terms in the potential for a scaling solution. The exponents in the
potential are described by the following seven vectors,
\ba
\bullet\; \underline\alpha_1&=&\frac{1}{\sqrt 21}\kappa(-3\sqrt 6,0,-4),\\\nonumber
\bullet\; \underline\alpha_2&=&\frac{1}{\sqrt 21}\kappa(-3\sqrt 6,\sqrt 7,3),\\\nonumber
\bullet\; \underline\alpha_3&=&\frac{1}{\sqrt 21}\kappa(-3\sqrt 6,-3\sqrt 7,3),\\\nonumber
\underline\alpha_4&=&\frac{1}{\sqrt 21}\kappa(-3\sqrt 6,-4\sqrt 7,-4),\\\nonumber
\underline\alpha_5&=&\frac{1}{\sqrt 21}\kappa(-3\sqrt 6,4\sqrt 7,-4),\\\nonumber
\underline\alpha_6&=&\frac{1}{\sqrt 21}\kappa(-3\sqrt 6,-2\sqrt 7,10),\\\nonumber
\underline\alpha_7&=&\frac{1}{\sqrt 21}\kappa(-7\sqrt 6,0,0).
\ea
In this case we find that a different structure appears. Whereas the SO(5)/SO(3)$_A$
case had directions in field space using the maximum number of terms in the potential
allowed by scaling, here we find that there can be at most one single term. 
By requiring that the negative terms $V_1$, $V_2$ and $V_3$ all decrease we find there
are regions where either $V_4$, $V_5$ or $V_6$ dominate
and each such region gives an exponent of $b=\sqrt{26/3}$ in (\ref{eqn:scalingPot}).
Thus SO(5)/SO(3)$_{A+B}$ can satisfy {\it strong scaling}.


\subsection{M(p,q,r)}
The coset SU(3)xSU(2)xU(1)/SU(2)xU(1)xU(1) leads to the potential given in (\ref{eqn:Mpqr}), which we
can describe using the following vectors,
\ba
\underline\alpha_1&=&\frac{1}{\sqrt 21}\kappa(-3\sqrt 6,2\sqrt 7,8\sqrt 2),\\\nonumber
\bullet\; \underline\alpha_2&=&\frac{1}{\sqrt 21}\kappa(-3\sqrt 6,-2\sqrt 7,\sqrt 2),\\\nonumber
\bullet\; \underline\alpha_3&=&\frac{1}{\sqrt 21}\kappa(-3\sqrt 6,\sqrt 7,\sqrt 2),\\\nonumber
\underline\alpha_4&=&\frac{1}{\sqrt 21}\kappa(-3\sqrt 6,-4\sqrt 7,8\sqrt 2),\\\nonumber
\underline\alpha_5&=&\frac{1}{\sqrt 21}\kappa(-7\sqrt 6,0,0).
\ea
When looking for {\it strong scaling} regions we discover that although there are four fields, at
most there are two that will be relevant. The terms that can dominate individually are $V_1$, $V_4$, $V_5$
for which we find the effective exponents $\sqrt{10}$, $\sqrt{14}$ and $\sqrt{14}$ respectively. There is
also a set of directions where both $V_1$ and $V_4$ dominate, the study of these directions will be included
in forthcoming work.

\subsection{N(p,q,r)}

The directions in the potential for N$^{pqr}$ are given by the following
vectors,
\ba
\underline\alpha_1&=&\frac{1}{\sqrt 21}\kappa(-3\sqrt 6,0, -4\sqrt 7,8\sqrt 2),\\\nonumber
\bullet\; \underline\alpha_2&=&\frac{1}{\sqrt 21}\kappa(-3\sqrt 6,0, -2\sqrt 7, \sqrt 2),\\\nonumber
\underline\alpha_3&=&\frac{1}{\sqrt 21}\kappa(-3\sqrt 6,-2\sqrt 21,-2\sqrt 7, \sqrt 2),\\\nonumber
\underline\alpha_4&=&\frac{1}{\sqrt 21}\kappa(-3\sqrt 6, 2\sqrt 21,-2\sqrt 7, \sqrt 2),\\\nonumber
\bullet\; \underline\alpha_5&=&\frac{1}{\sqrt 21}\kappa(-3\sqrt 6, -\sqrt 21,  \sqrt 7, \sqrt 2),\\\nonumber
\bullet\; \underline\alpha_6&=&\frac{1}{\sqrt 21}\kappa(-3\sqrt 6, \sqrt 21,  \sqrt 7, \sqrt 2),\\\nonumber
\underline\alpha_7&=&\frac{1}{\sqrt 21}\kappa(-3\sqrt 6,0         , 4\sqrt 7, \sqrt 2),\\\nonumber
\underline\alpha_8&=&\frac{1}{\sqrt 21}\kappa(-3\sqrt 6, 2\sqrt 21, 2\sqrt 7,8\sqrt 2),\\\nonumber
\underline\alpha_9&=&\frac{1}{\sqrt 21}\kappa(-3\sqrt 6,-2\sqrt 21, 2\sqrt 7,8\sqrt 2),\\\nonumber
\underline\alpha_{10}&=&\frac{1}{\sqrt 21}\kappa(-7\sqrt 6,0,0,0).
\ea
In this case each of the positive terms, $V_1$, $V_3$, $V_4$, $V_7$, $V_8$, $V_9$, $V_{10}$, 
in the potential can satisfy {\it strong scaling} individually.
In these cases the effective exponent $b$ in (\ref{eqn:scalingPot})
is $\sqrt{14}$, $\sqrt{8}$, $\sqrt{8}$, $\sqrt{8}$, $\sqrt{14}$, 
$\sqrt{14}$, $\sqrt{14}$, respectively.
We also find that there can be up to three dominant terms while still maintaining the {\it strong scaling}
criteria; such terms are given in table \ref{fig:scalingTerms}.


\subsection{Q(p,q,r)}

The directions in the Q$^{pqr}$ potential are described by the following
vectors,

\ba
\underline\alpha_1&=&\frac{1}{\sqrt 21}\kappa(-3\sqrt 6,\,2\sqrt 21,\,2\sqrt 7,\,8\sqrt 2),\\\nonumber
\bullet\; \underline\alpha_2&=&\frac{1}{\sqrt 21}\kappa(-3\sqrt 6,\,0,\, -2\sqrt 7,\, \sqrt 2),\\\nonumber
\bullet\; \underline\alpha_3&=&\frac{1}{\sqrt 21}\kappa(-3\sqrt 6,\, -\sqrt 21,\,\sqrt 7,\, \sqrt 2),\\\nonumber
\bullet\; \underline\alpha_4&=&\frac{1}{\sqrt 21}\kappa(-3\sqrt 6,\, \sqrt 21,\,\sqrt 7,\, \sqrt 2),\\\nonumber
\underline\alpha_5&=&\frac{1}{\sqrt 21}\kappa(-3\sqrt 6,\,-2\sqrt 21,\,2\sqrt 7,\,8\sqrt 2),\\\nonumber
\underline\alpha_6&=&\frac{1}{\sqrt 21}\kappa(-3\sqrt 6,\,0,\,-4\sqrt 7,\,8\sqrt 2),\\\nonumber
\underline\alpha_7&=&\frac{1}{\sqrt 21}\kappa(-7\sqrt 6,\,0,\,0,\,0).
\ea

\noindent
As in the N$^{pqr}$ case we find that each of the positive terms in the potential, $V_1$, $V_5$, $V_6$, $V_7$,
can satisfy {\it strong scaling} on their own, in each case we find that the effective exponent is $\sqrt{14}$.
We also find that there can be up to three terms in the potential which can dominate together over the rest,
these are shown in table \ref{fig:scalingTerms}.

\begin{table}
\center
\begin{tabular}{cc}
\hline
coset               & terms allowed\\
\hline
SO(5)/SO(3)$_A$     & \{2\},\{3\},\{4\},\{8\},\{9\},\{10\},\\
		    & \{2,3\},\{2,4\},\{2,9\},\{3,4\},\{3,8\},\{4,10\},\\
		    & \{2,3,4\},\{2,3,8\},\{2,3,9\},\{2,4,9\},\{2,4,10\},\{3,4,8\},\{3,4,10\},\\
		    & \{2,3,4,8\},\{2,3,4,9\},\{2,3,4,10\}\\
SO(5)/SO(3)$_{A+B}$ &  \{4\},\{5\},\{6\}\\
SO(5)/SO(3)$_{max}$ & \{2\} \\
M${}^{pqr}$         & \{1\},\{4\},\{5\},\{1,4\}\\
N${}^{pqr}$         & \{1\},\{3\},\{4\},\{7\},\{8\},\{9\},\{10\},\\
		    & \{1,3\},\{1,4\},\{1,8\},\{1,9\},\{3,9\},\{3,10\},\{4,8\},\{4,10\},\{7,8\},\{7,9\},\{7,10\},\{8,9\},\\
		    & \{1,3,9\},\{1,4,8\},\{1,8,9\},\{7,8,9\},\{1,3,8,9\},\{1,4,8,9\},\{1,7,8,9\}\\

Q${}^{pqr}$         & \{1\},\{5\},\{6\},\{7\},\{1,5\},\{1,6\},\{5,6\}\{1,5,6\}.\\
\hline
\end{tabular}
\caption{
Terms allowed in {\it strong scaling} directions for the cosets of 11d supergravity.
}
\label{fig:scalingTerms}
\end{table}

\section{Concluding remarks}
\label{sec:conclusions}

In this paper we have given an explicit construction of the consistent dimensional reduction of eleven
dimension supergravity over the homogeneous spaces classified in \cite{Castellani:1983yg}.
By including a Freund-Rubin four-form flux
we have seen that this leads to an effective theory in four dimensions that consists of Einstein
gravity coupled to a series of moduli fields, and that these fields experience a potential which
is composed of a series of exponentials. Motivated by the large body of material which studies
gravity+scalars with exponential potentials we studied the possibility of scaling solutions in
this setup. However, for exact scaling solutions to exist there can be at most the same number
of terms in the potential as there are fields; if this is violated then one finds an over constrained
algebraic system for which there is no solution. In all the cosets of the classification we find that
there are more terms in the potential than there are moduli fields. While this rules out the existence
of exact scaling solutions, one still has the possibility that there are regions in field space where
only a subset of terms are relevant, thereby allowing approximate scaling solutions. Having given a set
of criteria for such regions to exist we showed that it is possible to get approximate scaling in each
of the coset examples. 
The cases which we analysed explicitly were only the ones where there is a single
dominant term, and we explicitly gave the exponent of the effective potential. We also noted that there
are examples where more than one term could dominate,
giving the possibility of assisted behaviour in these directions; we shall leave the detailed study of such
examples for future work.

Aside from the Freund-Rubin flux that we included, there is also the possibility that one may include
internal flux. This internal flux may take the form of a combination of axions coming from 
three-form potentials as $C_3\sim\phi(x)c_3(y)$, and non-exact flux,
i.e. a four form flux that cannot be written globally as d$c$. In such a scheme, the consistent truncation
would require us to take fluxes that matched the symmetries of the coset involved and would lead
to a finite, manageable set of scalar fields. This would then lead to a four dimensional effective theory
containing axions and geometrical moduli. Both of these types fields are to be expected in such dimensional
reduction schemes and it remains to be seen how the presence of such axions in coset models affects the
dynamics.

\vspace{1cm}
\noindent
{\large\bf Acknowledgements} The authors would like to thank Beatriz de Carlos, Mark Hindmarsh, Thomas House
and Eran Palti for useful discussions. Both JLPK and PMS are supported by PPARC.

\vskip 1cm
\appendix{\noindent\Large \bf Appendices}
\renewcommand{\theequation}{\Alph{section}.\arabic{equation}}
\setcounter{equation}{0}

\section{Review of invariant objects on a coset}
\label{AppCosets}
Decompose the Lie algebra as ${\cal G}={\cal H}\oplus {\cal K}$,
where ${\cal H}\subset {\cal G}$ is a subgroup of ${\cal G}$.
We consider only compact groups, so the coset space $G/H$ is reductive
\cite{Duff:1986hr} i.e.
\ba
[{\cal H}_a,{\cal H}_b]&=&{\cal H}_cf^c_{\;\;ab},\\\nonumber
[{\cal H}_a,{\cal K}_i]&=&{\cal K}_jf^j_{\;\;ai},\\\nonumber
[{\cal K}_i,{\cal K}_j]&=&{\cal H}_af^a_{\;\;ij}+{\cal K}_kf^k_{\;\;ij}.
\ea
We use $i,j,...$ indices to label elements of ${\cal K}$ and $a,b,...$ to label
elements of ${\cal H}$. 
Now consider the left-invariant form on $G/H$, $\Theta$, which we expand by
introducing the one-forms $e^a$ and $e^i$.
\ba
\Theta=L^{-1}dL={\cal H}_ae^a+{\cal K}_ie^i.
\ea
To find the algebra of these one-forms consider
\ba
d\Theta=dL^{-1}\wedge dL=-L^{-1}dL\wedge L^{-1}dL=-\Theta\wedge\Theta
\ea
which then gives
\ba
de^a&=&-\half f^a_{\;\;bc}e^b\wedge e^c-\half f^a_{\;\;ij}e^i\wedge e^j\\\nonumber
de^i&=&-\half f^i_{\;\;jk}e^j\wedge e^k-f^i_{\;\;aj}e^a\wedge e^j
\ea
We may use these left-invariant one-forms to construct a homogeneous, G-invariant
metric on $G/H$,
\ba
ds^2_{G/H}&=&g_{ij}e^i\otimes e^j.
\ea
Homogeneity requires the parameters $g_{ij}$ to be independent of the co-ordinates
on $G/H$ and $G$-invariance requires \cite{Mueller-Hoissen:1987cq}
\ba
\label{eqn:metricGinv}
g_{kj}f^k_{\;ia}+g_{ik}f^k_{\;ja}=0.
\ea
This relation gets generalised for objects with more indices, such as form fields
\ba
\label{eqn:tensorGinv}
T_{ki_2i_3...}f^k_{\;i_1a}+T_{i_1ki_3...}f^k_{\;i_2a}+T_{i_1i_2k...}f^k_{\;i_3a}+...=0.
\ea
Note that for $G$ to be unimodular then $f^I_{\;\;IJ}=0$ and for $H$ to be unimodular
then $f^a_{\;\;ab}=0$ and for $G/H$ being reductive then $f^a_{\;\;bi}=0$, all of
which show that $f^i_{\;\;ij}=0$ and $f^i_{\;\;ia}=0$.

\section{Reducing the Ricci scalar}
\label{AppReduceRicci}
We choose a higher dimensional metric to consist of a spacetime part and
an internal coset part according to
\ba
ds^2&=&e^{2\psi(x)}ds^2_{(1,d-1)}+g_{ij}(x)e^i\otimes e^j,\\\nonumber
    &=&e^{2\psi(x)}\eta_{\mu\nu}e^\mu\otimes e^\nu+g_{ij}(x)e^i\otimes e^j\\\nonumber
    &=&\hat g_{\hat\mu\hat\nu}e^{\hat\mu}\otimes e^{\hat\nu},
\ea
with the co-ordinates on spacetime being represented by $x$ and those on the
coset by $y$, $\psi(x)$ represents a freedom to choose the spacetime co-ordinates.
In the following we shall analyse this space using the frame $e^{\hat\mu}=(e^\mu,e^i)$,
note that this is not an orthonormal frame.
In order to find the connection one-forms, $\omega^{\hat\mu}_{\;\;\hat\nu}$, we
need to solve
\ba
d\hat g_{\hat\mu\hat\nu}-\omega^{\hat\rho}_{\;\;\hat\mu}g_{\hat\rho\hat\nu}
                        -\omega^{\hat\rho}_{\;\;\hat\nu}g_{\hat\mu\hat\rho}&=&0\\\nonumber
de^{\hat\mu}+\omega^{\hat\mu}_{\;\;\hat\nu}\wedge e^{\hat\nu}&=&0,
\ea
and the curvature two-forms follow from
\ba
\hat R^{\hat\mu}_{\;\;\hat\nu}&=&d\omega^{\hat\mu}_{\;\;\hat\nu}+
                            \omega^{\hat\mu}_{\;\;\hat\rho}\wedge \omega^{\hat\rho}_{\;\;\hat\nu}.
\ea
We find that the Ricci tensor is given by
\ba
\label{eqn:scaleFactorEOM}
\hat {\cal R}_{\mu\nu}&=&{\cal R}_{(d)\mu\nu}
  -(d-2)\nabla_\mu\nabla_\nu\psi
  -\eta_{\mu\nu}\nabla_\rho\nabla^\rho\psi
  -(d-2)\eta_{\mu\nu}\nabla_\rho\psi\nabla^\rho\psi
  +(d-2)\nabla_\mu\psi\nabla_\nu\psi\\\nonumber
  &~&-\frac{1}{4}\nabla_\mu g^{ij}\nabla_\nu g_{ij}
  -\half g^{ij}\nabla_\mu\nabla_\nu g_{ij}
  +\half g^{ij}\left(\nabla_\mu g_{ij}\nabla_\nu\psi+\nabla_\nu g_{ij}\nabla_\mu\psi  \right)
  -\half \eta_{\mu\nu}g^{ij}\nabla_\rho g_{ij}\nabla^\rho \psi\\
\label{eqn:mixedRicci}
\hat {\cal R}_{\mu j}&=&-\half g^{kl}\nabla_\mu g_{km}f^m_{\;\;lj}\\\nonumber
\label{eqn:shapeModuliEOM}
\hat {\cal R}_{ij}&=&\tilde{\cal R}_{ij}
  +e^{-2\psi}\left(  \half g^{kl}\nabla_\mu g_{ik}\nabla^\mu g_{jl}
                    -\half\nabla_\mu\nabla^\mu g_{ij}
                    +\frac{1}{4}g_{kl}\nabla_\mu g^{kl}\nabla^\mu g_{ij}
                    -\half (d-2)\nabla_\mu \psi\nabla^\mu g_{ij}\right).
\ea
In deriving this we have used the fact that compact Lie groups are unimodular,
giving $f^I_{IJ}=0$ \cite{Scherk:1979zr,Cvetic:2003jy,Cho:1975sf}.
$\tilde{\cal R}_{ij}$ denotes the curvature of the coset space, treating the
$g_{ij}$ as constant and the covariant derivatives, $\nabla_\mu$ are for
the metric $ds^2_{(1,d-1)}$ with their indices raised by $\eta^{\mu\nu}$.
Given the Ricci curvatures above we can see one of the issues related to
the consistency of truncation, namely that there is nothing to source 
${\cal R}_{\mu j}$ and so it must vanish by the 11D equations of motion.
For the cases we consider, we find that this term does vanish.

We may now trace the above to find the following Ricci scalar
\ba
\hat {\cal R}&=&{\cal R}_{G/H}
           +e^{-2\psi}[ {\cal R}_{(d)}-2(d-1)\nabla^2\psi-(d-1)(d-2)\nabla_\mu\psi\nabla^\mu \psi
                      -g^{ij}\nabla^2 g_{ij}\\\nonumber
      &~&             -\frac{3}{4}\nabla_\mu g^{ij}\nabla_\mu g_{ij}
                      -(d-2)g^{ij}\nabla_\mu g^{ij}\nabla^\mu\psi
                      -\frac{1}{4}g^{ij}\nabla_\mu g_{ij}g^{kl}\nabla^\mu g_{kl}].
\ea

\noindent Making use of the gauge freedom we choose
\ba
\label{eqn:gaugeChoice}
e^{(d-2)\psi}\sqrt{g_{ij}}=1
\ea
showing that the physical volume of the internal space is given by
\ba
V_{phys}&=&V_{G/H}e^{(2-d)\psi}.
\ea
This gauge choice enables us to write
\ba
\label{eqn:RicciWithGauge}
\hat {\cal R}_{\mu\nu}&=&{\cal R}_{(d)\mu\nu}+\frac{1}{2(d-2)}g^{ij}\nabla_\sigma\nabla^\sigma g_{ij}\eta_{\mu\nu}
                    +\frac{1}{4}\nabla_\mu g^{ij}\nabla_\nu g_{ij}
                    +\frac{1}{2(d-2)}\eta_{\mu\nu}\nabla_\sigma g^{ij}\nabla^\sigma g_{ij}\\\nonumber
           &~&      -\frac{1}{4(d-2)}g^{ij}\nabla_{\mu}g_{ij} g^{kl}\nabla_\nu g_{kl}\\\nonumber
       &=&{\cal R}_{(d)\mu\nu}+\frac{1}{4}\nabla_\mu g^{ij}\nabla_\nu g_{ij}
            -\eta_{\mu\nu}\nabla^2\psi-(d-2)\nabla_\mu\psi\nabla_\nu\psi\\\nonumber
\hat {\cal R}_{ij}&=&\tilde{\cal R}_{ij}+\half e^{-2\psi}\left[ g^{kl}\nabla_\mu g_{ik}\nabla^\mu g_{jl}
                                                          -\nabla_\mu\nabla^\mu g_{ij}\right]\\\nonumber
\hat {\cal R}&=&e^{-2\psi}[ {\cal R}_{(d)}-2(d-1)\nabla^2\psi-g^{ij}\nabla^2 g_{ij}
                       -\frac{3}{4}\nabla_\mu g^{ij}\nabla_\mu g_{ij}
                       -\frac{1}{4(d-2)}g^{ij}\nabla_\mu g_{ij}g^{kl}\nabla^\mu g_{kl}]\\\label{eqn:RicciScalarWithGauge}
           &~&+{\cal R}_{G/H}.
\ea

\section{Coset SO(5)/SO(3)$_A$}
\label{AppSO5SO3A}
\subsection{The invariant metric}
The Lie algebra of SO(n) may be written in terms of one-forms, $\Lambda_{IJ}$
as
\ba
{\rm d}\Lambda_{IJ}&=&-2\Lambda_{IK}\wedge\Lambda_{KJ}
\ea
where $\Lambda_{IJ}=-\Lambda_{JI}$ and $I,J,...=0,1,...n-1$.
Taking $N=5$ we can introduce a different basis, $\lambda_i$, $\rho_i$ and $\Pi_a$
\ba
\lambda_i&=&-2\left( \Lambda_{0i}+\half\epsilon_{ijk}\Lambda_{jk}\right)\\\nonumber
\rho_i&=&2\left( \Lambda_{0i}-\half\epsilon_{ijk}\Lambda_{jk}\right)\\\nonumber
\Pi_p&=&2\Lambda_{p4}.
\ea
In this basis we find that the $\lambda_i$ and $\rho_i$ form commuting so(3) 
sub-algebras, labelled so(5)$_A$ and so(5)$_B$ respectively. 
Given these sub-algebras we may construct the coset SO(5)/SO(3)$_A$.
In order to find a $G$ invariant metric we apply 
the techniques of appendix \ref{AppCosets} to $g_{ab}$ using (\ref{eqn:tensorGinv}) 
and discover that the most general metric on the coset, consistent with its symmetries
is
\ba
g_{ab}&=&\left( \begin{array}{cccccccc}
         \alpha & \phi   & \theta & 0      & 0      & 0      & 0      \\
         \phi   & \beta  & \psi   & 0      & 0      & 0      & 0      \\
         \theta & \psi   & \gamma & 0      & 0      & 0      & 0      \\
         0      & 0      & 0      & \delta & 0      & 0      & 0      \\
         0      & 0      & 0      & 0      & \delta & 0      & 0      \\
         0      & 0      & 0      & 0      & 0      & \delta & 0      \\
         0      & 0      & 0      & 0      & 0      & 0      & \delta
         \end{array} \right),
\ea
so there are seven allowed parameters.
\subsection{The Ricci tensor}
In order to simplify the analysis of the dynamics we restrict the most general metric
to the diagonal case, $g_{ab}={\rm diag}(\alpha,\beta,\gamma,\delta,\delta,\delta,\delta)$
and find that the Ricci tensor is given by
\ba
\label{SO3ARicci}
{\cal R}_{11} &=&
    \frac{\alpha^2}{2 \beta \gamma} -\frac{\beta}{2 \gamma}
	-\frac{\gamma}{2 \beta} + \frac{\alpha^2}{\delta^2} + 1,\\\nonumber
{\cal R}_{22} &=&
    \frac{\beta^2}{2 \alpha \gamma} -\frac{\alpha}{2 \gamma}
	-\frac{\gamma}{2 \alpha} + \frac{\beta^2}{\delta^2} + 1,\\\nonumber
{\cal R}_{33} &=&
    \frac{\gamma^2}{2 \alpha \beta} -\frac{\alpha}{2 \beta}
	-\frac{\beta}{2 \alpha} + \frac{\gamma}{\delta^2} + 1,\\\nonumber
{\cal R}_{44} =  {\cal R}_{55} =  {\cal R}_{66} =  {\cal R}_{77} &=&
    -\frac{\alpha}{2 \delta} -\frac{\beta}{2 \delta}
    - \frac{\gamma}{2 \delta} + 3.
\ea
When considering the dynamics one discovers that there are special points in the
moduli space given by Einstein metrics, corresponding to stationary points of
the effective potential. This condition states that the Ricci curvature is proportional
to the metric, we find two such points given by \cite{Castellani:1983yg}
\ba
\alpha&=&\beta=\gamma=\delta\\
{\cal R}_{ab}&=&\frac{3}{2\alpha}g_{ab}
\ea
and
\ba
\label{eqn:so3AEinstein}
\alpha&=&\beta=\gamma=\delta/5\\
{\cal R}_{ab}&=&\frac{27}{50\alpha}g_{ab}.
\ea

\section{Coset SO(5)/SO(3)$_{A+B}$}
\label{AppSO5SO3A+B}
\subsection{The invariant metric}
One may construct another SO(3) subgroup of SO(5) based on the diagonal part
of SO(3)$_A\times$SO(3)$_B$, SO(3)$_{A+B}$. To see this we introduce a new basis
of one-forms,
\ba
\sigma_i&=&- \epsilon_{ijk} \Lambda_{jk} = \half(\lambda_i+\rho_i),\\\nonumber
\tau_i&=& 2 \Lambda_{0i} = - \half(\lambda_{i}-\rho_{i}),\\\nonumber
\tau_4&=& 2 \Lambda_{04} = \Pi_0,\\\nonumber
\tau_5&=& 2 \Lambda_{14} = \Pi_1,\\\nonumber
\tau_6&=& 2 \Lambda_{24} = \Pi_2,\\\nonumber
\tau_7&=& 2 \Lambda_{34} = \Pi_3.
\ea
In this basis we find that the $\sigma_i$ form a commuting so(3) 
sub-algebra, labelled so(5)$_{A+B}$. 
In order to find a $G$ invariant metric we apply (\ref{eqn:tensorGinv}) to $g_{ab}$
and discover that the most general metric on the coset consistent with its symmetries
is
\ba
g_{ab}&=&\left( \begin{array}{cccccccc}
         \alpha & 0      & 0      & 0      & \delta & 0      & 0      \\
         0      & \alpha & 0      & 0      & 0      & \delta & 0      \\
         0      & 0      & \alpha & 0      & 0      & 0      & \delta \\
         0      & 0      & 0      & \beta  & 0      & 0      & 0      \\
         \delta & 0      & 0      & 0      & \gamma & 0      & 0      \\
         0      & \delta & 0      & 0      & 0      & \gamma & 0      \\
         0      & 0      & \delta & 0      & 0      & 0      & \gamma 
         \end{array} \right),
\ea
and so there are four allowed parameters.

\subsection{The Ricci scalar}
Again we restrict to a diagonal metric given by
$g_{ab}={\rm diag}(\alpha,\alpha,\alpha,\beta,\gamma,\gamma,\gamma)$
and find that the Ricci tensor has the following components
\ba
\label{SO3A+BRicci}
{\cal R}_{11} =  {\cal R}_{22} =  {\cal R}_{33} &=&
    \frac{\alpha^2}{2 \beta \gamma} -\frac{\beta}{2 \gamma}
    -\frac{\gamma}{2 \beta} + 3,\\\nonumber
{\cal R}_{44}&=& 
    -\frac{3 \alpha}{2 \gamma} + \frac{3 \beta^2}{2 \alpha \gamma}
    -\frac{3 \gamma}{2 \alpha} + 3,\\\nonumber
{\cal R}_{55} =  {\cal R}_{66} =  {\cal R}_{77} &=&
    -\frac{\alpha}{2 \beta} -\frac{\beta}{2 \alpha}
	+ \frac{\gamma}{2 \alpha \beta} + 3.
\ea
We may also find the points in moduli space for which such metrics satisfy the
Einstein condition, we find \cite{Castellani:1983yg}
\ba
\alpha&=&\gamma=\frac{2}{3}\beta\\
{\cal R}_{ab}&=&\frac{9}{4\alpha}g_{ab}.
\ea

\section{Coset SO(5)/SO(3)$_{MAX}$}
\label{AppSO5SO3MAX}
\subsection{The invariant metric}
The final SO(3) subgroup of SO(5) is the maximal subgroup, which we can see by
introducing the following basis of one-forms
\ba
\sigma_1&=& \frac{1}{\sqrt{5}} (\lambda_1 + \sqrt{3} \Pi_2),\\\nonumber
\sigma_2&=& \frac{1}{\sqrt{5}} (\lambda_2 + \sqrt{3} \Pi_1),\\\nonumber
\sigma_3&=& \frac{1}{2\sqrt{5}} (\lambda_3 - 3 \rho_3),\\\nonumber
\tau_1&=& - \frac{1}{4\sqrt{5}} (3\lambda_1 -2\sqrt{3}\Pi_2 -5\rho_1),\\\nonumber
\tau_2&=& - \frac{1}{4\sqrt{5}} (3\lambda_2 -2\sqrt{3}\Pi_1 -5\rho_2),\\\nonumber
\tau_3&=& \frac{1}{2\sqrt{5}} (3\lambda_3 + \rho_3),\\\nonumber
\tau_4&=& - \frac{1}{4} (\sqrt{3}\lambda_1 - 2\Pi_2 + \sqrt{3}\rho_1),\\\nonumber
\tau_5&=& \, \frac{1}{4} (\sqrt{3}\lambda_2 - 2\Pi_1 + \sqrt{3}\rho_2),\\\nonumber
\tau_6&=& \Pi_0,\\\nonumber
\tau_7&=& \Pi_3.
\ea
In this basis we find that the $\sigma_i$ form a commuting so(3) 
sub-algebra, labelled so(5)$_{MAX}$, we shall now consider
the coset SO(5)/SO(3)$_{MAX}$.
In order to find a $G$ invariant metric we apply (\ref{eqn:tensorGinv}) to $g_{ab}$
and discover that the most general metric on the coset consistent with its symmetries
is
\ba
g_{ab}&=&\left( \begin{array}{cccccccc}
         \alpha & 0      & 0      & 0      & 0      & 0      & 0      \\
         0      & \alpha & 0      & 0      & 0      & 0      & 0      \\
         0      & 0      & \alpha & 0      & 0      & 0      & 0      \\
         0      & 0      & 0      & \alpha & 0      & 0      & 0      \\
         0      & 0      & 0      & 0      & \alpha & 0      & 0      \\
         0      & 0      & 0      & 0      & 0      & \alpha & 0      \\
         0      & 0      & 0      & 0      & 0      & 0      & \alpha 
         \end{array} \right),
\ea
so there is only one allowed parameter and the metric is already diagonal. One finds
that this metric is already an Einstein metric, with the Ricci curvature being given by
\ba
\label{SO3MAXRicci}
{\cal R}_{ab} = \frac{27}{10\alpha}g_{ab}.
\ea

\section{Coset M${}^{pqr}$ = SU(3) x SU(2) x U(1) / U(1) x U(1)}
\label{AppMpqr}
\subsection{The invariant metric}
The generators of the Lie algebra of $M^{pqr}$ may be written as,
\ba
T^1 &=& -\half i \lambda_4 \qquad
    T^2 = -\half i \lambda_5 \qquad
    T^3 = -\half i \lambda_6\\\nonumber
T^4 &=& -\half i \lambda_7 \qquad
    T^5 = -\half i \sigma_1 \qquad
    T^6 = -\half i \sigma_2\\\nonumber
T^7 &=& -\frac{i}{2\zeta} \left(
	    2 r y + \sqrt{3} p \lambda_8 + q \sigma_3
	\right)\\\nonumber
T^8 &=& -\half i \lambda_1 \qquad
    T^9 = -\half i \lambda_2 \qquad
    T^{10} = -\half i \lambda_3\\\nonumber
T^{11} &=& \frac{i}{\sqrt{2} \zeta\eta} \left(
	    (3p^2 + q^2) y - \sqrt{3} p r \lambda_8 - q r \sigma_3
	\right)\\\nonumber
T^{12} &=& - \frac{i}{2\eta} \left(\sqrt{3} p \sigma_3 - q \lambda_8 \right)
\ea
where $\sigma_i$ with $i \in \{1, 2, 3\}$ are the Pauli matrices
with $ [\sigma_i, \sigma_j] = 2 i \, \epsilon_{ijk} \, \sigma_k$, and the $\lambda_i$
with $i \in \{1..8\}$ are the Gell-Mann $SU(3)$ matrices normalised
such that $ [\lambda_i, \lambda_j] = 2 i \, f_{ijk} \, \lambda_k$ with
the totally antisymmetric $\epsilon$ and $f$ given by,

\ba
\epsilon_{123} &=& 1\\\nonumber
f_{123} &=& 1\\\nonumber
f_{147} &=& - f_{156} = f_{246} = f_{257} = f_{345} = - f_{367} = \half\\\nonumber
f_{458} &=& f_{678} = \frac{\sqrt{3}}{2},
\ea

\noindent and $y$ is the generator of the separate U(1) factor.  The
three primes p, q and r characterise the embedding of the U(1) x
U(1) subgroup.  For convenience we define the following quantities,

\be
\zeta = \sqrt{3p^2 + q^2 + 2 r^2} \qquad \eta = \sqrt{3p^2 + q^2}.
\ee

In order to find a $G$ invariant metric we apply the techniques of
appendix \ref{AppCosets} to $g_{ab}$ using (\ref{eqn:tensorGinv})
and discover that the most general metric on the coset, consistent
with its symmetries is

\ba
\label{Mmetric}
g_{ab}&=&\left( \begin{array}{cccccccc}
         \alpha & 0      & 0      & 0      & 0      & 0      & 0      \\
         0      & \alpha & 0      & 0      & 0      & 0      & 0      \\
         0      & 0      & \alpha & 0      & 0      & 0      & 0      \\
         0      & 0      & 0      & \alpha & 0      & 0      & 0      \\
         0      & 0      & 0      & 0      & \beta  & 0      & 0      \\
         0      & 0      & 0      & 0      & 0      & \beta  & 0      \\
         0      & 0      & 0      & 0      & 0      & 0      & \gamma
         \end{array} \right),
\ea
so there are three allowed parameters.

\subsection{The Ricci tensor}
Taking the metric to be, $g_{ab}={\rm
diag}(\alpha,\alpha,\alpha,\alpha,\beta,\beta,\gamma)$
and find that the Ricci tensor is given by
\ba
\label{MRicci}
{\cal R}_{11} = {\cal R}_{22} =
{\cal R}_{33} = {\cal R}_{44} &=&
    \frac{3}{2} - \frac{9}{8}\, \frac{p^2}{{\zeta }^2}\,
	\frac{\gamma }{\alpha},\\\nonumber
{\cal R}_{55} = {\cal R}_{66} &=&
    1 - \frac{1}{2}\, \frac{q^2}{{\zeta }^2}\,
	\frac{\gamma }{\beta },\\\nonumber
{\cal R}_{77} &=&
    \frac{\gamma^2}{2\, \zeta^2} \, \left(
	\frac{9}{2}\, \frac{p^2}{\alpha^2} + \frac{q^2}{\beta^2}
    \right).
\ea

\section{Coset N${}^{pqr}$ = SU(3) x U(1) / U(1) x U(1)}
\label{AppNpqr}
\subsection{The invariant metric}
The generators of the Lie algebra of $N^{pqr}$ may be written as,
\ba
T^1 &=& -\half i \lambda_1 \qquad
    T^2 = -\half i \lambda_2 \qquad
    T^3 = -\half i \lambda_4\\\nonumber
T^4 &=& -\half i \lambda_5 \qquad
    T^5 = -\half i \lambda_6 \qquad
    T^6 = -\half i \lambda_7\\\nonumber
T^7 &=& -\frac{i}{2\zeta} \left(
	    2 r y + q \lambda_3 + \sqrt{3} p \lambda_8
	\right)\\\nonumber
T^8 &=& \frac{i}{\sqrt{2} \zeta\eta} \left(
	    (3p^2 + q^2) y - q r \lambda_3 - \sqrt{3} p r \lambda_8
	\right)\\\nonumber
T^9 &=& - \frac{i}{2\eta} \left(\sqrt{3} p \lambda_3 - q \lambda_8 \right)
\ea
where $\lambda_i$ with $i \in \{1..8\}$ are the Gell-Mann $SU(3)$
matrices normalised such that $ [\lambda_i, \lambda_j] = 2 i \, f_{ijk} \,
\lambda_k$ with the totally antisymmetric $f$ given by,

\ba
f_{123} &=& 1\\\nonumber
f_{147} &=& - f_{156} = f_{246} = f_{257} = f_{345} = - f_{367} = \half\\\nonumber
f_{458} &=& f_{678} = \frac{\sqrt{3}}{2},
\ea

\noindent and $y$ is the generator of the separate U(1) factor.  The
three primes p, q and r characterise the embedding of the U(1) x
U(1) subgroup.  For convenience we define the following quantities,

\be
\zeta = \sqrt{3p^2 + q^2 + 2 r^2} \qquad \eta = \sqrt{3p^2 + q^2}.
\ee

In order to find a $G$ invariant metric we apply the techniques of
appendix \ref{AppCosets} to $g_{ab}$ using (\ref{eqn:tensorGinv})
and discover that the most general metric on the coset, consistent
with its symmetries is

\ba
\label{Nmetric}
g_{ab}&=&\left( \begin{array}{cccccccc}
         \alpha & 0      & 0      & 0      & 0      & 0      & 0      \\
         0      & \alpha & 0      & 0      & 0      & 0      & 0      \\
         0      & 0      & \beta  & 0      & 0      & 0      & 0      \\
         0      & 0      & 0      & \beta  & 0      & 0      & 0      \\
         0      & 0      & 0      & 0      & \gamma & 0      & 0      \\
         0      & 0      & 0      & 0      & 0      & \gamma & 0      \\
         0      & 0      & 0      & 0      & 0      & 0      & \delta
         \end{array} \right),
\ea
so there are four allowed parameters.

\subsection{The Ricci tensor}
Taking the metric to be, $g_{ab}={\rm
diag}(\alpha,\alpha,\beta,\beta,\gamma,\gamma,\delta)$
and find that the Ricci tensor is given by
\ba
\label{NRicci}
{\cal R}_{11} = {\cal R}_{22} &=&
    \frac{3}{2} - \frac{q^2}{2\,\zeta^2} \, \frac{\delta}{\alpha}
    + \frac{1}{4\,\beta\,\gamma}\left(
	    \alpha^2 - \beta^2 - \gamma^2 \right),\\\nonumber
{\cal R}_{33} = {\cal R}_{44} &=&
    \frac{3}{2} - \frac{(3\,p+q)^2}{8\,\zeta^2} \, \frac{\delta}{\beta}
    - \frac{1}{4\,\alpha\,\gamma}\left(
	    \alpha^2 - \beta^2 + \gamma^2 \right),\\\nonumber
{\cal R}_{55} = {\cal R}_{66} &=&
    \frac{3}{2} - \frac{(3\,p-q)^2}{8\,\zeta^2} \, \frac{\delta}{\gamma}
    - \frac{1}{4\,\alpha\,\beta}\left(
	    \alpha^2 + \beta^2 - \gamma^2 \right),\\\nonumber
{\cal R}_{77} &=&
    \frac{\delta^2}{2\, \zeta^2}\, \left(
	\frac{q^2}{\alpha^2}
	+ \frac{(3\, p + q)^2}{4\, \beta^2}
	+ \frac{(3\, p - q)^2}{4\, \gamma^2}
    \right).
\ea

\section{Coset Q${}^{pqr}$ = SU(2) x SU(2) x SU(2) / U(1) x U(1)}
\label{AppQpqr}
\subsection{The invariant metric}
The generators of the Lie algebra of $Q^{pqr}$ may be written as,
\ba
T^1 &=& -\frac{i}{1}\, \sigma_1, \qquad
    T^2 = -\frac{i}{2}\, \sigma_2, \qquad
    T^3 = -\frac{i}{2}\, {\tilde\sigma}_1,\\\nonumber
T^4 &=& -\frac{i}{2}\, {\tilde\sigma }_2, \qquad
    T^5 = -\frac{i}{2}\, {\tilde\sigma }_1, \qquad
    T^6 = -\frac{i}{2}\, {\tilde\sigma }_2, \\\nonumber
T^7 &=& -\frac{i}{2\, \zeta}\, \left( 
	    p\, {\sigma }_3
	    + q\, {\tilde\sigma}_3
	    + r\, {\hat\sigma }_3
	\right), \\\nonumber
T^8 &=& -\frac{i }{2\, \zeta \, \eta}\, \left(
	    p\, r\, \sigma_3
	    + q\, r\, {\tilde\sigma }_3
	    - (p^2 + q^2) \, {\hat\sigma }_3
	\right), \\\nonumber
T^9 &=& \phantom{-} \frac{i }{2\, \eta}\, \left(
            p\, {\tilde\sigma }_3
	    - q\, {\sigma }_3
	\right)
\ea
where $\sigma_i, \tilde\sigma_i, \hat\sigma_i$ with $i \in \{1, 2, 3\}$ are
three sets of mutually commuting Pauli matrices.

\noindent The three primes p, q and r characterise the embedding
of the U(1) x U(1) subgroup.  For convenience we define the following
quantities,

\be
\zeta = \sqrt{p^2 + q^2 + r^2} \qquad \eta = \sqrt{p^2 + q^2}.
\ee

In order to find a $G$ invariant metric we apply the techniques of
appendix \ref{AppCosets} to $g_{ab}$ using (\ref{eqn:tensorGinv})
and discover that the most general metric on the coset, consistent
with its symmetries is

\ba
g_{ab}&=&\left( \begin{array}{cccccccc}
         \alpha & 0      & 0      & 0      & 0      & 0      & 0      \\
         0      & \alpha & 0      & 0      & 0      & 0      & 0      \\
         0      & 0      & \beta  & 0      & 0      & 0      & 0      \\
         0      & 0      & 0      & \beta  & 0      & 0      & 0      \\
         0      & 0      & 0      & 0      & \gamma & 0      & 0      \\
         0      & 0      & 0      & 0      & 0      & \gamma & 0      \\
         0      & 0      & 0      & 0      & 0      & 0      & \delta
         \end{array} \right),
\ea
so there are four allowed parameters.

\subsection{The Ricci tensor}
Taking the metric to be $g_{ab}={\rm
diag}(\alpha,\alpha,\beta,\beta,\gamma,\gamma,\delta)$
we find that the Ricci tensor is given by
\ba
\label{QRicci}
{\cal R}_{11} = {\cal R}_{22} &=&
    1 - \frac{1}{2}\, \frac{p^2}{\zeta^2}\, \frac{\delta}{\alpha},\\\nonumber
{\cal R}_{33} = {\cal R}_{44} &=&
    1 - \frac{1}{2}\, \frac{q^2}{\zeta^2}\, \frac{\delta}{\beta},\\\nonumber
{\cal R}_{55} = {\cal R}_{66} &=&
    1 - \frac{1}{2}\, \frac{r^2}{\zeta^2}\, \frac{\delta}{\gamma},\\\nonumber
{\cal R}_{77} &=&
    \frac{\delta^2}{2\, \zeta^2}\, \left(
	\frac{p^2}{\alpha^2} + \frac{q^2}{\beta^2} + \frac{r^2}{\gamma^2}
    \right).
\ea


\end{document}